\documentclass[floats,floatfix,amssymb,prd,twocolumn,superscriptaddress,nofootinbib]{revtex4-1}

\usepackage{subcaption}
\usepackage{ragged2e}
\DeclareCaptionJustification{justified}{\justifying}
\makeatletter
\newcommand{\subsetsim}{\mathrel{\mathpalette\subset@sim\relax}}
\newcommand{\subset@sim}[2]{%
  \vtop{\offinterlineskip\m@th
    \ialign{\hfil##\cr
      $#1\subset$\cr\noalign{\kern0.5pt}\scalebox{0.9}{$#1\sim$}\cr
    }%
  }%
}
\makeatother

\usepackage{amssymb,amsmath,verbatim,mathtools,needspace,enumitem,etoolbox,graphicx,physics,microtype,afterpage,bm}
\usepackage[dvipsnames, usenames]{xcolor}
\definecolor{linkcolor}{rgb}{0.0,0.3,0.5}
\usepackage{booktabs}
\usepackage[unicode, colorlinks=true, linkcolor=linkcolor, citecolor=linkcolor, filecolor=linkcolor,urlcolor=linkcolor, pdfusetitle]{hyperref}
\usepackage[all]{hypcap}
\usepackage[T1]{fontenc}
\usepackage[utf8]{inputenc}
\usepackage{tabularx}
\usepackage{float}
\renewcommand{\arraystretch}{1.4}

\usepackage{multirow}
\usepackage{pifont}
\usepackage{lmodern}

\usepackage{multirow}

\allowdisplaybreaks
\usepackage{tikz}
\usepackage{color}
\usepackage{framed}
\usepackage{hyperref}
\hypersetup{colorlinks, citecolor=bluscuro, linkcolor=black, urlcolor=bluscuro}
\definecolor{rossos}{cmyk}{0,1,1,0.55}
\definecolor{bluscuro}{rgb}{0.15, 0.2, .85}
\definecolor{bluchiaro}{cmyk}{1,.3,0.,0.1}
\definecolor{ForestGreen}{rgb}{0.13, 0.55, 0.13}
\definecolor{azure}{rgb}{0.0, 0.5, 1.0}
 
\usepackage{slashed}

\newcommand{\apjl}{"Astrophys. J. Lett."}

\newcommand{\aap}{"Astronomy and Astrophysics"}

\def\nn{\nonumber}

\def\bea{\begin{eqnarray}}
\def\eea{\end{eqnarray}}

\def\d{{\mathrm{d}}}

\newcommand{\bs}{\begin{subequations}}
\newcommand{\es}{\end{subequations}}

\newcommand{\be}{\begin{equation}}
\newcommand{\ee}{\end{equation}}
\renewcommand{\d}{{\rm d}}

\def\lsim{\mathrel{\rlap{\lower4pt\hbox{\hskip0.5pt$\sim$}}
    \raise1pt\hbox{$<$}}}         
\def\gsim{\mathrel{\rlap{\lower4pt\hbox{\hskip0.5pt$\sim$}}
    \raise1pt\hbox{$>$}}}         

\newcommand{\unmezzo}{\frac{1}{2}}

\DeclareMathOperator\arctanh{arctanh}

\renewcommand{\d}{{\rm d}}
\newcommand{\eff}{{\rm eff}}

\makeatletter
\def\l@subsubsection#1#2{}
\makeatother

\newcommand{\sapienza}{Dipartimento di Fisica, Sapienza Università 
	di Roma, Piazzale Aldo Moro 5, 00185, Roma, Italy}
\newcommand{\infn}{INFN, Sezione di Roma, Piazzale Aldo Moro 2, 00185, Roma, Italy}

\begin{document}
\title{
Fermion Soliton Stars
}

\begin{abstract}
A real scalar field coupled to a fermion via a Yukawa term can evade no-go theorems preventing solitonic solutions. For the first time, we study this model within General Relativity without approximations, finding static and spherically symmetric solutions that describe fermion soliton stars. The Yukawa coupling provides an effective mass for the fermion, which is key to the existence of self-gravitating relativistic solutions.
We systematically study this novel family of solutions and present their mass-radius diagram and maximum compactness, which is close to (but smaller than) that of the corresponding Schwarzschild photon sphere. Finally, we discuss the ranges of the parameters of the fundamental theory in which the latter might have interesting astrophysical implications, including compact (sub)solar and supermassive fermion soliton stars for a standard gas of degenerate neutrons and electrons, respectively.
\end{abstract}

\author{Loris Del Grosso}
\affiliation{\sapienza}
\affiliation{\infn}

\author{Gabriele Franciolini}
\affiliation{\sapienza}
\affiliation{\infn}

\author{Paolo Pani}
\affiliation{\sapienza}
\affiliation{\infn}

\author{Alfredo Urbano}
\affiliation{\sapienza}
\affiliation{\infn}

\date{\today}
\maketitle

{
  \hypersetup{linkcolor=black}
  \tableofcontents
}

\section{Introduction}\label{sec:intro}

Solitonic solutions play a crucial role in many field theories, in particular in General Relativity.
In the context of the latter, starting from Wheeler's influential idea of geons~\cite{Wheeler:1955zz}, considerable attention has been devoted to find minimal models allowing for self-gravitating solitonic solutions~\cite{Herdeiro:2015waa}.
The prototypical example is that of boson stars~\cite{Kaup:1968zz,Ruffini:1969qy,Colpi:1986ye} (and of their Newtonian analog, Q-balls~\cite{Coleman:1985ki}), which are self-gravitating solutions to the Einstein-Klein-Gordon theory with a complex and massive scalar field (see~\cite{Jetzer:1991jr,Schunck:2003kk,Liebling:2012fv} for some reviews).
If the scalar field is real, no-go theorems prevent the existence of solitonic solutions for very generic classes of scalar potential~\cite{Derrick:1964ww,Herdeiro:2019oqp}. Indeed, the Einstein-Klein-Gordon theory contains time-dependent solutions known as oscillatons which, however, decay in time~\cite{Seidel:1991zh}.

Solitonic configurations were constructed also with non-zero spin fields. A prototypical example is given by Dirac stars~\cite{PhysRevD.59.104020}, which are solutions of the Einstein-Dirac equations with two neutral fermions. An example of self-gravitating configurations supported
by a complex spin-1 field is provided by Proca stars~\cite{Brito:2015pxa}. More complex theories, in which both fermion and vector fields are present, were also studied (see e.g.~\cite{PhysRevD.101.024023}).

About 40 years ago, Lee and Pang proposed a model in which a real scalar field with a false-vacuum potential is coupled to a massive fermion via a Yukawa term~\cite{Lee:1986tr}. Working in a thin-wall limit in which the scalar field is a step function, for certain parameters of the model they obtained approximated solutions describing \emph{fermion soliton stars}.

The scope of this paper is twofold. On the one hand we show that fermion soliton stars exist in this model also beyond the thin-wall approximation, and we build exact static solutions within General Relativity. On the other hand, we elucidate some key properties of the model, in particular the role of the effective fermion mass provided by the Yukawa coupling. Then, we explore the model systematically, presenting mass-radius diagrams and the maximum compactness of fermion soliton stars for various choices of the parameters, showing that in this model a standard gas of degenerate neutrons (resp.\ electrons) can support stable (sub)solar (resp.\ supermassive) fermion soliton stars with compactness comparable to that of ordinary neutron stars.
This is particularly intriguing in light of the fact that some of the detected LIGO-Virgo events (e.g., GW190814~\cite{LIGOScientific:2020zkf} and GW190521~\cite{LIGOScientific:2020iuh}, in the lower and upper mass gaps, respectively) might not fit naturally within the standard astrophysical formation scenarios for black holes and neutron stars and are compatible with more exotic origins (e.g.,~\cite{CalderonBustillo:2020fyi}).
Our analysis paves the way for a detailed study of the phenomenology of fermion soliton stars as a motivated model of exotic compact objects~\cite{Cardoso:2019rvt}.
Finally, in Appendix~\ref{appendix}, we explore the connection of the model to a very peculiar scalar-tensor theory.

We use the signature $(-,+,+,+)$ for the metric, adopt natural units ($\hbar = c = 1$) and define the Planck mass through $G = m_p^{-2}$.

\section{Setup}

We consider a theory in which Einstein gravity
is minimally coupled to a real scalar field $\phi$
and a fermion field $\psi$.
The action can be written as~\cite{Lee:1986tr}
\begin{align}\label{theory_fund}
    S = \int \d^4 x \sqrt{-g}
    \Big[
    &\frac{R}{16\pi G} - \unmezzo \partial^\mu \phi \partial_\mu \phi - U(\phi)
    \nonumber \\
    &+\bar{\psi}(i\gamma^\mu D_\mu - m_f)\psi + f\phi \bar{\psi} \psi\Big],
\end{align}
where the scalar potential is
\begin{equation}\label{our_potential}
		U (\phi) = \unmezzo \mu^2 \phi^2 \Big(1- \frac{\phi}{\phi_0}\Big)^2,
\end{equation}
and features two degenerate minima at $\phi=0$ and $\phi=\phi_0$. The constant $\mu$ (resp. $m_f$) is the mass of the scalar (resp. fermion). The Yukawa interaction\footnote{See also Ref.~\cite{Garani:2022quc} for a recent work on a condensed dark matter in a model with a Yukawa coupling between a fermion and a scalar particle.} is controlled by the coupling $f$. The fermionic field has a $U(1)$ global symmetry which ensures the conservation of the fermion number $N$.
It should be noted that Eq.~\eqref{theory_fund} describes the action of a local field theory and, therefore, we expect that all physics derived from it will naturally respect causality conditions (that, on the contrary, could be violated in the absence of such underlying formulation).
Also, we point out that the matter Lagrangian in Eq.~\eqref{theory_fund} describes a renormalizable field theory; this is in contrast to the widely used model describing solitonic boson stars~\cite{Friedberg:1986tq,Palenzuela:2017kcg,Bezares:2022obu,Boskovic:2021nfs} in which the scalar potential is non-renormalizable and field values should not exceed the limit of validity of the corresponding effective field theory.
The covariant derivative $D_\mu$ in Eq.~\eqref{theory_fund} takes into account the spin connection of the fermionic field. 

From the quadratic terms in the fermion Lagrangian, it is useful to define an effective mass,
\begin{equation}
	m_\eff = m_f - f\phi.
\end{equation}
We will focus on scenarios in which the fermion becomes effectively massless 
(i.e. $m_\eff = 0$)
when the scalar field sits on the
second degenerate vacuum, $\phi=\phi_0$. This condition implies
fixing
\begin{equation}\label{eq:f_fixing}
	f  =\frac{m_f}{\phi_0}.
\end{equation}

As we shall discuss, we are mostly interested in configurations where the scalar field makes a transition between the false\footnote{Although the minima at $\phi=0$ and $\phi=\phi_0$ are degenerate, we shall call them true and false vacuum, respectively, having in mind the generalization in which the potential $U(\phi)$ can be nondegenerate, i.e. $U(\phi_0)\neq U(0)$ (see Fig.~\ref{fig:inverted_potential}).} vacuum ($\phi \approx\phi_0$) to the true vacuum ($\phi\approx0$).\footnote{
Recently, Ref.~\cite{Hong:2020est} studied a related model in which dark fermions are trapped inside the false
vacuum during a first-order cosmological phase transition, 
subsequently forming compact macroscopic ``Fermi-balls'', which are dark matter candidates and can collapse to primordial black holes~\cite{Kawana:2021tde}.} 

\subsection{Thomas-Fermi approximation}

The description of a fermionic field in Eq.~\eqref{theory_fund} requires treating the quantization of spin-1/2 particles in curved spacetime. In particular, one should deal with the problem of finding the ground state of an ensemble of $N$ fermions in curved spacetime (see e.g.~\cite{PhysRevD.59.104020,PhysRevD.104.046024}). However, in the macroscopic limit $N\gg1$, 
it is convenient to adopt a mean-field approach, which in this context is called the Thomas-Fermi approximation\footnote{We point the interested reader to Appendix A of Ref.~\cite{Lee:1986tr} for a complete derivation of the Thomas-Fermi approximation in curved spacetime, while here we summarise the main properties.}
. The latter relies on the assumption that the gravitational and scalar fields 
are slowly varying functions
with respect to the fermion dynamics. Consequently, they do not interact directly with the (microscopic) fermionic field $\psi$, but with average macroscopic quantities.
In practice, one can divide the entire three-space into small domains which are much larger than the de~Broglie wavelength of the typical fermion, but sufficiently small that the gravitational and scalar fields are approximately constant inside each domain. 
Then, every domain is filled with a degenerate (i.e. the temperature is much smaller than the chemical potential) Fermi gas, in such a way that the Fermi distribution is approximated by a step function, $n_k = \theta(k_{\rm F} - k)$, 
where $k_{\rm F}(x^\mu)$
is the  Fermi momentum observed in the appropriate local frame. 

The energy density of the fermion gas reads
\begin{equation}\label{fermion_energy}
	W = \frac{2}{(2\pi)^3} \int_0^{k_{\rm F}} \d^3 k \, \epsilon_k,
\end{equation}
where $\epsilon_k = \sqrt{k^2 + m_\eff^2}$.
Notice that $W=W(x^\mu)$ through the spacetime dependence of $k_{\rm F}$ and $m_\eff$.
In an analogous way, we obtain the fermion gas pressure $P$ and the scalar density $S = \langle \bar{\psi}\psi \rangle$ as
\begin{align}\label{fermion_pressure}
	P &= \frac{2}{(2\pi)^3} \int_0^{k_{\rm F}} \d^3 k \hspace{0.1cm} \frac{k^2}{3\epsilon_k},
\\
\label{fermion_density}
S &=  \frac{2}{(2\pi)^3} \int_0^{k_{\rm F}} \d^3 k \hspace{0.1cm} \frac{m_\eff}{\epsilon_k}.
\end{align}
It it easy to show that these quantities satisfy the identity
\begin{equation}\label{identityWPF}
	W - 3P = m_\eff S.
\end{equation}

In the Thomas-Fermi approximation, the fermions enter Einstein's equations as a perfect fluid 
characterized by an energy-momentum tensor of the form
\begin{equation}
	T^{[f]}_{\mu\nu} = (W+P)u_\mu u_\nu + Pg_{\mu\nu},
\end{equation}
while they also enter the scalar field equation through the scalar density $S$.
Indeed, by varying the action in Eq.~\eqref{theory_fund} with respect to $\phi$, we obtain a source term of the form $\approx f\bar{\psi}\psi$. 
Within the Thomas-Fermi approximation, this becomes 
\begin{equation}
f\bar\psi \psi   
\rightarrow	f \langle \bar{\psi}\psi \rangle \equiv fS,
\end{equation} 
which is consistent with the fact that, in the fluid description, the scalar field equation couples to fermions through 
a term proportional to the trace $(T^{[f]})^{\mu}_\mu = -W + 3P$. 

\subsubsection{Equations of motion}

It is now possible to write down the equations of motion for our theory in covariant form
\begin{align}\label{equations_covariant_form}
    &G_{\mu\nu} = 8\pi G\, T_{\mu\nu}\nn, \\ 
    &\Box \phi - \frac{\partial U}{\partial \phi} + fS = 0,
\end{align}
where
\begin{equation}
    T_{\mu\nu} = -2\Big(\frac{\partial \mathcal{L}_\phi}{\partial g^{\mu\nu}}-\unmezzo g_{\mu\nu} \mathcal{L}_\phi\Big) + T^{[f]}_{\mu\nu}, 
\end{equation}
in which $\mathcal{L}_\phi$ is the Lagrangian density of the scalar field. 
In order to close the system, we need an equation describing the behavior of $k_{\rm F}$. 
This is obtained by minimizing the energy of the fermion gas at fixed number of fermions~\cite{Lee:1986tr}.

From now on, for simplicity, we will consider spherically symmetric equilibrium configurations,
whose background metric can be expressed as
\begin{equation}\label{eq:general_spacetime}
\d s^2 = 
-e^{2u(\rho)} \d t^2 
+ e^{2v(\rho)}\d \rho^2 
+ \rho^2 (\d \theta^2 + \sin^2\theta \d \varphi^2),
\end{equation}
in terms of two real metric functions $u(\rho)$ and $v(\rho)$.
Furthermore, we will assume that the scalar field in its equilibrium configuration is also static and spherically symmetric, $\phi(t,\rho,\theta,\varphi) = \phi(\rho)$.
Being the spacetime static and spherically symmetric, $k_{\rm F} = k_{\rm F}(\rho)$ can only be a function of the radial coordinate.
\vspace{0.2cm}

\subsubsection{Fermi momentum equation}
In the Thomas-Fermi approximation the fermion gas energy can be written as~\cite{Lee:1986tr}
\begin{equation}
	E_f = 
	4\pi \int \d\rho\, \rho^2 \, e^{u(\rho) + v(\rho)} \,W,
\end{equation}
while the number of fermions is
\begin{equation}
N = 
\frac{4}{3 \pi}
\int \d \rho\,  \rho^2 e^{v(\rho)}
k_{\rm F}^3(\rho).
\end{equation}
To enforce a constant number of fermions, we introduce the Lagrangian multiplier $\omega_{\rm F}$ and define the functional 
\begin{equation}
	E_f'[k_{\rm F}] = E_f[k_{\rm F}] - \omega_{\rm F}\Big(N[k_{\rm F}] - N_{\rm fixed}\Big),
\end{equation}
which is minimized by imposing 
\begin{equation}
	\frac{\delta E_f'[k_{\rm F}]}{\delta k_{\rm F}(\rho)} = 0.
\end{equation}
This directly brings us to the condition
\begin{equation}\label{minimize_condition}
\epsilon_{\rm F} = e^{-u}\omega_{\rm F}, 
\end{equation}
where $\epsilon_{\rm F} = \epsilon_{k_{\rm F}}$ is the Fermi energy. Thus, $\omega_F$ coincides with the Fermi energy in flat spacetime while it acquires a redshift factor otherwise.
Since $\epsilon_F = \sqrt{k_F^2 + m_{\rm eff}^2}$, Eq.~\eqref{minimize_condition} in turn gives 
\begin{equation}\label{eq:kFermi}
 k_{\rm F}^2(\rho) = \omega_{\rm F}^2e^{-2u(\rho)} - (m_f - f\phi(\rho))^2\,.
\end{equation}
\subsection{Dimensionless equations of motion and boundary conditions}

In order to simplify the numerical integrations, as well as physical intuition, it is convenient 
writing the field equations in terms of dimensionless quantities.
To this end, we define 
\begin{equation}\label{dimensionless_variables}
x  = \frac{k_{\rm F}}{m_f}, \qquad
y  = \frac{\phi}{\phi_0}, \qquad
r = \rho \mu .  
\end{equation}
Therefore, the potential and kinetic terms become
\begin{align}\label{}
U & = \mu^2 \phi_0^2 \left [ \unmezzo y^2 (1 - y)^2 \right ]
\equiv  \mu^2 \phi_0^2 \tilde{U}(y) ,
\nonumber 
\\
V &
= \mu^2 \phi_0^2 \left [\unmezzo e^{-2 v(r)} (\partial_r y)^2 \right ]
\equiv \mu^2 \phi_0^2 \tilde{V}(y),
\end{align}
while Eqs.~\eqref{fermion_energy}-\eqref{fermion_density} can be computed analytically as
\begin{widetext}
\begin{subequations}
\begin{align}\label{eq:WPSexplicit}
	&	W = 
	 \frac{2}{(2\pi)^3}\int_{0}^{k_{\rm F}(\rho)} \d^3 k \, \sqrt{k^2 + (m_f - f\phi(\rho))^2}
	 =
\frac{m_\eff^4}{8\pi^2} 
\left[
s\sqrt{1+s^2} (1 + 2s^2) - \log(s + \sqrt{s^2 + 1})
\right] 
\equiv 
m_f^4 \tilde{W}(x,y) ,
	 \\
	 \label{eq:Pexplicit}
& P = 
\frac{2}{(2\pi)^3}\int_{0}^{k_{\rm F}(\rho)} 
\frac{\d^3 k \,  k^2}{3 \sqrt{k^2 + (m_f - f\phi(\rho))^2}} 
=
\frac{m_\eff^4}{8\pi^2} 
\left[s\left (\frac{2}{3}s^2 - 1 \right )\sqrt{1 + s^2} +  \log(s + \sqrt{s^2 + 1})\right]
\equiv  m_f^4 \tilde{P}(x,y) ,
\\
\label{eq:Sexplicit}
&	S = 
	\frac{2}{(2\pi)^3}\int_{0}^{k_{\rm F}(\rho)} \d^3 k \, \frac{m_f - f\phi(\rho)}{\sqrt{k^2 + (m_f - f\phi(\rho))^2}}
	=
\frac{m_\eff^3}{2\pi^2} 
\left[s\sqrt{1 + s^2} -  \log(s + \sqrt{s^2 + 1})\right] 
	\equiv 
	m_f^3 \tilde{S}(x,y),
\end{align}
\end{subequations}
\end{widetext}
where $\tilde{W}, \tilde{P}, \tilde{S}$ are dimensionless quantities and
we introduced $s \equiv {x}/(1-y)$ for convenience.
Remarkably, these expressions are the same as in the standard case of a minimally coupled degenerate gas with the substitution $m_f \to m_{\rm eff}$.

As we shall discuss in Appendix~\ref{appendix}, this property will be important when comparing this model to a scalar-tensor theory. 
Note that the massless limit, $m_{\rm eff}\to0$, should be taken carefully so as not all the dependence on $m_{\rm eff}$ is expressed in the dimensional prefactor. By performing the first integrals in Eqs.~\eqref{eq:WPSexplicit}-\eqref{eq:Sexplicit} in the $m_{\rm eff}\to0$ limit, we obtain $W=P/3$, as expected for an ultrarelativistic degenerate gas.

It is convenient to further introduce the dimensionless combination of parameters
\begin{align}\label{dimensionless_parameters}
	\Lambda &= \frac{\sqrt{8\pi}\phi_0}{m_p}, 
	\qquad 
	\eta  = \frac{m_f}{\mu^{1/2} \phi_0^{1/2}}.
\end{align}
Finally, the field equations (i.e.\ the Einstein-Klein-Gordon equations with the addition of the Fermi momentum equation) take the compact form
\begin{align}\label{fund_sistema_dimensionlesse_dimensionlesse}
& e^{-2v}-1-2  e^{-2v} r\partial_r v = -\Lambda^2 r^2  \left [ \eta^4 \tilde{W} + \tilde{U} +  \tilde{V} \right],
\nn 
\\
&
e^{-2 v} - 1 + 2  e^{- 2v} r\partial_r u =\Lambda^2 r^2  \left [\eta^4 \tilde{P} - \tilde{U} +  \tilde{V}\ \right],
\nn 
\\
&
e^{-2v}\Big[  \partial_r ^2 y +  \Big(\partial_r u - \partial_r v + \frac{2}{r}\Big)\partial_r y \Big] 
		= \frac{\partial \tilde{U}}{\partial y} - \eta^4\tilde{S} ,
		\nn 
		\\
& 
x^2 
= \tilde{\omega}_{\rm F}^2 e^{-2 u (r)} - (1-y)^2,
\end{align}
where $\tilde U$, $\tilde V$, $\tilde P$, $\tilde W$, and $\tilde S$ depend on $x$, $y$, and $r$, and we also introduced 
$\tilde{\omega}_{\rm F} = {\omega_{\rm F}}/{m_f}$.
Static and spherically symmetric configurations in the model~\eqref{theory_fund} are solutions to the above system of ordinary differential equations.
For clarity, we summarize the relevant parameters in Table~\ref{Tb:parameters}.

\subsubsection{Absence of $\phi={\rm const}$ solutions}

Note that, because $U=0=dU/d\phi$ in both degenerate vacua, it is natural to first check what happens when $\phi=\phi_0={\rm const}$ or if $\phi=0$. The former case (i.e., $y(\rho)=1$) is an exact solution of the scalar equation and reduces Einstein's equations to those of gravity coupled to a degenerate gas of {\rm massless} (since $m_\eff (\phi_0) =0$) fermions. 
In this case, self-gravitating solutions do not have a finite radius~\cite{Shapiro:1983du}.
On the other hand, due to the Yukawa coupling, in the presence of a fermion gas $\phi=0$ is not a solution to the scalar field equation. 

Thus, self-gravitating solutions to this model must have a nonvanishing scalar-field profile. In particular, we will search for solutions that (approximately) interpolate between these two vacuum states.

{
\renewcommand{\arraystretch}{1.4}
\setlength{\tabcolsep}{4pt}
\begin{table}[t]
\caption{List of the model parameters, the fermion soliton star parameters, and the dimensionless quantities adopted to express the system of equations in compact form. Due to the condition in  Eq.~\eqref{eq:f_fixing}, in our case only three model parameters are independent.}
\begin{tabularx}{1 \columnwidth}{|X|c|}
\hline
\hline
\multicolumn{2}{|c|}{ \it Model parameters }    
\\
\hline
\hline
$\mu$   & Scalar field mass
\\
\hline
$\phi_0$  & VEV of the false vacuum
\\
\hline
$m_f$  & Fermion mass
\\
\hline
$f$ & Yukawa coupling
\\
\hline
\hline
\multicolumn{2}{|c|}{\it Solution parameters (boundary conditions)}    
\\
\hline
\hline
$P_c$ & Fermion central pressure
\\
\hline
$\epsilon= 1-\phi/\phi_0$ & Central scalar field displacement
\\
\hline
\hline
\multicolumn{2}{|c|}{\it Dimensionless parameters/variables}    
\\
\hline
\hline
$\Lambda=\sqrt{8\pi}\phi_0/m_p$ & Dimensionless VEV of the false vacuum
\\
\hline
$\eta=m_f/\sqrt{\mu\phi_0}$ & Scale ratio
\\
\hline
\hline
$x=k_{\rm F}/m_f$ & Fermi momentum 
\\
\hline
$y=\phi/\phi_0$ & Scalar field
\\
\hline
$r=\rho\mu$ & Rescaled radius
\\
\hline
\hline
\end{tabularx}
\label{Tb:parameters}
\end{table}
}

\subsubsection{Boundary conditions at $\rho = 0$}

Regularity at the center of the star ($
\rho=0$) imposes the following boundary conditions 
\begin{align}
\label{initial_conditions_dimensionlessi}
v(r = 0) & = 0, 
\qquad\ \, \, 
u(r = 0) = 0, 
\nn \\
y(r = 0) & = 1 - \epsilon, 
\qquad 	
\partial_r y(0) = 0, 
\nn \\
\tilde{P}(r = 0) &= \tilde{P}_c,
\end{align}
where $\epsilon>0$ will be fixed numerically through a shooting procedure in order to obtain asymptotic flatness. 

The central value of the pressure $\tilde{P}_c$ is fixed in terms of $\tilde{\omega}_{\rm F}$ and $\epsilon$ through the relation
\begin{align}\label{Pcvsomegafepsilon}
    \tilde{P}_c = \frac{1}{24\pi^2}\Bigg(\tilde{\omega}_{\rm F} \sqrt{\tilde{\omega}_{\rm F}^2 - \epsilon^2}&(2\tilde{\omega}_{\rm F}^2-5\epsilon^2)\nonumber\\
    &+3\epsilon^4 \arctanh{\sqrt{1-\frac{\epsilon^2}{\tilde{\omega}_{\rm F}^2}}}\Bigg),
\end{align}
obtained computing Eq.~\eqref{eq:Pexplicit} in $\rho = 0$.
In practice, in a large region of the parameter space one obtains $\epsilon\ll1$. In this limit, Eq.~\eqref{Pcvsomegafepsilon} reduces to $\tilde{P}_c \approx \tilde{\omega}_{\rm F}^4 / 12 \pi^2$.

%
%

Finally, since a shift $u(\rho)\rightarrow u(\rho) + {\rm const}$ in Eq.~\eqref{fund_sistema_dimensionlesse_dimensionlesse} merely corresponds to a shift of the fermionic central pressure, we have imposed $u(\rho = 0) = 0$ without loss of generality.

\subsubsection{Definitions of mass, radius, and compactness}
We define the mass of the object as \begin{equation}
	M = \frac{m(\rho \to +\infty)}{G},
\end{equation}
where the function $m(\rho)$ is related to the metric coefficient 
$v(\rho)$ by $e^{2v(\rho)} = 1- 2m(\rho)/\rho$, 
and can be interpreted as the mass energy enclosed within the radius $\rho$.
In terms of the dimensionless variables introduced in Eq.~\eqref{dimensionless_variables}, it is convenient to define $\tilde{m}(r) = \mu m(\rho)$. Thus, one obtains
\begin{equation}
	\dfrac{\mu M}{m_p^2} = \tilde{m}(r).
	\label{eq:defmass}
\end{equation}
Notice that, in the asymptotic limit $r \to \infty$, Eq.~\eqref{eq:defmass} becomes independent of the radius. 

Typically, the radius of a star is defined 
as the value of the radial coordinate 
at the point where pressure drops to zero.
As we shall discuss, in our case the fermion soliton stars will be characterized by a lack of a sharp boundary. 
Analogously to the case of boson stars~\cite{Liebling:2012fv}, 
one can define an effective radius $R$ within which
$99\%$ of the total mass is contained. (As later discussed, we shall also define the location $R_f$ where only the pressure of the fermion gas vanishes.)
Finally, we can define the compactness of the star as ${G M}/{R}$.
%

\section{Some preliminary theoretical considerations}\label{section_someptc}
Before solving the full set of field equations numerically, in this section we provide some theoretical considerations that might be useful to get a physical intuition of the model. 

\subsection{On the crucial role of fermions for the existence of solitonic stars}

\subsubsection{Classical mechanics analogy}

In order to understand why the presence of fermions in this theory plays a crucial role for the existence of stationary solutions, it is useful to study a classical mechanics analogy for the dynamics of the scalar field~\cite{Coleman:1985ki}. 

For the moment we consider flat spacetime. Furthermore, we start by ignoring the fermions (we will relax this assumption later on). The set of Eqs.~\eqref{fund_sistema_dimensionlesse_dimensionlesse} drastically simplifies to a single field equation 
\begin{equation}\label{eq:scalar_field_no_gravity}
	\partial_\rho ^2 \phi +  \frac{2}{\rho}\partial_\rho \phi  - \frac{\partial U}{\partial \phi} = 0.
\end{equation}
To make the notation more evocative of a one-dimensional mechanical system, we rename
\begin{equation}\label{analogy}
\rho  \rightarrow t ,
\quad 
\phi(\rho) \rightarrow \phi(t),
\quad 
\hat{U} := -U,
\end{equation}
in such a way that the equation of motion becomes
\begin{equation}
	\phi''(t) = -\frac{\partial \hat{U}}{\partial \phi} - \frac{2}{t} \phi'(t),
\end{equation}
which describes the one-dimensional motion of a particle with coordinate $\phi(t)$ in the
presence of an \textit{inverted} potential,
$\hat{U}$, and a velocity-dependent dissipative force, $-({2}/t) \phi'(t)$. 
Within this analogy, the boundary (or initial) conditions~\eqref{initial_conditions_dimensionlessi} simply become
\begin{equation}\label{initial_conditions_no_gravity}
\phi(t = 0) = \phi_0 - \delta\phi ,
\quad
\phi'(t = 0) = 0,
\end{equation}
where $\phi_0$ is the position of the false vacuum and $\delta \phi=\epsilon \phi_0$. As we impose zero velocity at $t = 0$, the initial energy 
is $E({0}) = \hat{U}(\phi_0 - \delta\phi)$.
The energy $E(t)$ of the particle at a time $t$ is obtained by subtracting the work done by the friction:
\begin{equation}
	E(t) - E({0}) = L(t),
\end{equation}
where
\begin{equation}
	L(t) = -2\int_{0}^{t} \d t' \,\frac{\dot{\phi}^2(t')}{t'}.
\end{equation}
Note that, owing to the initial conditions, this integral is regular at $t = 0$.
 On the other hand, the existence of a solution with asymptotically zero energy requires
the particle to arrive with zero velocity at $\phi = 0$ for $t \to +\infty$. 
Therefore, we impose $E(t\to {\infty})=0$. 
As the total energy loss due to friction is $L(t\to {\infty})$, the latter condition means

\begin{equation}\label{eq:tosolve}
E(0)  = - L(t\to {\infty}) 
\end{equation}
that is 
\begin{equation}\label{eq:tosolve2}
\hat{U}(\phi_0 - \delta\phi)=2\int_{0}^{\infty}\d t'  \frac{\dot{\phi}^2(t')}{t'}   \,.
\end{equation}

\begin{figure}[t]
	\centering
	\includegraphics[width=1\linewidth]{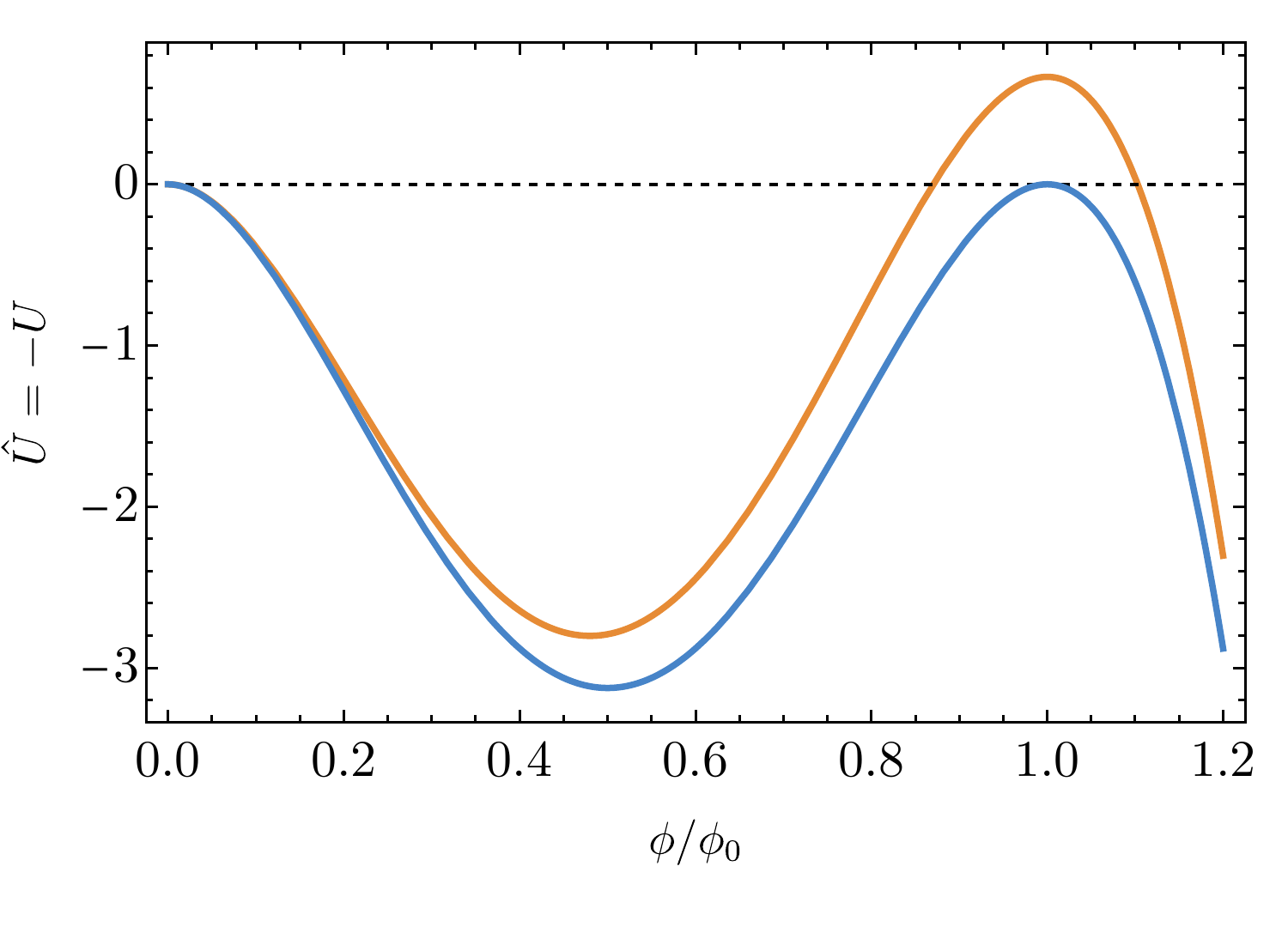}
	\caption{Inverted potential with degeneracy  (blue line, our case) and without degeneracy between vacua (orange line).}
	\label{fig:inverted_potential}
\end{figure}

This equation can be interpreted as an equation for $\delta\phi$ in order to allow for the existence of a ``bounce'' solution\footnote{A bounce solution is the one reaching asymptotically the true vacuum with zero energy, after having "bounced" at the minimum of the inverted potential.}. 
One can demonstrate the existence of such a solution heuristically. 
Let us first consider a slightly modified version of the inverted potential without degeneracy
(orange plot in Fig.~\ref{fig:inverted_potential}). 
Obviously, if the motion starts exactly at $\phi_0$ with zero velocity, the particle would remain at rest. 
However, if we start on the left of the maximum the particle will roll down, bounce, and eventually climb the leftmost hill shown in Fig.~\ref{fig:inverted_potential}. Now, if the dynamics starts \textit{too far from} $\phi_0$ (still on the left of the maximum), with zero initial velocity it might not have enough energy to reach the zero-energy point at $\phi = 0$. Similarly, if the dynamics starts \textit{too close} to $\phi_0$, the particle might reach $\phi=0$ with positive energy and \textit{overcome} the hill rolling up to $\phi \to -\infty$. 
By continuity, there must exist a unique point such that the total energy loss due to friction compensates the initial gap of energy with respect to the energy of $\phi = 0$.

However, by applying the same argument to our \textit{degenerate} case (blue curve in Fig.~\ref{fig:inverted_potential}), it is easy to see that there is no solution to Eq.~\eqref{eq:tosolve2} \footnote{At least if we look for a solution in which the scalar field does the transition at a finite time.}. 
This is because the energy loss due to friction is nonzero, so the particle will never reach $\phi=0$ and is doomed to roll back in the potential eventually oscillating around the minimum of $\hat{U}$. 
This shows that, in the degenerate case considered in this work, a simple scalar model does not allow for bounce solutions in flat spacetime. 
	
If we now reintroduce fermions in the theory, 
the scalar field equation reads (still in flat spacetime)
\begin{equation}
	\phi''(t) = -\frac{\partial \hat{U}}{\partial \phi} - \frac{2}{t} \phi'(t) - fS.
\end{equation}
Since $S \geq 0$, the fermions act with a force pushing our particle toward the origin, potentially giving the right kick to allow the particle reaching $\phi=0$ asymptotically. 
As we shall see, this also requires $S=0$ (i.e., no fermions) around the origin, in order for the particle  to reach a stationary configuration at $\phi=0$.

This simple analogy shows how the presence of the fermions is fundamental
as it allows the solution to exist.
In the following section we will show how this is realized in the full theory
which includes gravitational effects.
Furthermore, we will show that, in certain regions of the parameter space, relativistic effects are in fact crucial for the existence of the solution, since the latter requires a minimum fermionic pressure to exist.
	
\subsubsection{Evading the no-go theorem for solitons}
The above conclusions, deduced from our simple heuristic picture, holds also in the context of General Relativity. 
Indeed, without fermions in the system of Eqs.~\eqref{fund_sistema_dimensionlesse_dimensionlesse}, 
and since our potential~\eqref{our_potential} is nonnegative, a general theorem proves that no axially symmetric and stationary solitons (that is asymptotically flat, localized and everywhere regular solutions) can exist~\cite{Derrick:1964ww,Herdeiro:2019oqp}.

However, the presence of fermions evades one of the hypotheses of the theorem.
As we will show, in this case stationary solitons generically exist also for a \emph{real} scalar field (at variance with the case of boson stars, that require complex scalars) and for a wide choice of the parameters.

\subsection{Scaling of the physical quantities in the $\mu R\gg1$ regime}
Assuming $\mu R\gg 1$, it is possible to derive an analytical scaling for various physical quantities, as originally derived in Ref.~\cite{LEE1992251} and similar in spirit to Landau's original computation for ordinary neutron stars (see, e.g.,~\cite{Shapiro:1983du}). 

It is instructive to consider \eqref{theory_fund} in the absence of gravity. As already pointed out, that the theory has a conserved (additive) quantum number $N$, brought by the fermion field $\psi$. 
Being $\mu R \gg 1$, the real scalar field solution is well approximated by a stiff Fermi function~\cite{Lee:1986tr},\cite{LEE1992251}
\begin{equation}\label{eq:scalar_field_profile}
	\phi(\rho) \approx \frac{\phi_0}{1 + e^{\mu(\rho - R)}}.
\end{equation}
The definition of $k_{\rm F}$ is nothing but Eq.~\eqref{eq:kFermi} with $u = 0$ (since we work in absence of gravity)
\begin{equation}\label{eq:kFermiFlatSpace}
 k_{\rm F}^2(\rho) = \omega_{\rm F}^2 - (m_f - f\phi(\rho))^2\,.
\end{equation}
Because of Eq.~\eqref{eq:scalar_field_profile}, the Fermi momentum is nearly fixed to the constant value $\omega_{\rm F}$ for $\rho \lesssim R$, and for $\rho \approx R$ it goes to zero stiffly. Therefore, the field $\psi$ is approximately confined within the sphere of radius $R$. We assume that the quanta of $\psi$ are noninteracting, massless 
and described by Fermi statistics at zero temperature. Thus, we obtain the standard relation for the particle density
\begin{equation}
	n = \frac{\rm \#particles}{\rm  unit.volume} 
	= \frac{2}{8\pi^3} \int_{0}^{k_{\rm F}} 4\pi k^2 \d k 
	= \frac{\omega_{\rm F}^3}{3\pi^2}.
\end{equation}
Since $k_{\rm F} \simeq \omega_{\rm F} = {\rm const}$, 
the total number of particles is
\begin{equation}
N = n \int_{0}^{R} 4\pi \rho^2  \d \rho 
= \frac{4}{9\pi}(R\omega_{\rm F})^3.
\end{equation}
The fermion energy is
\begin{equation}
E_f =  \int_{0}^{R} 4\pi \rho^2\d \rho \hspace{0.1cm} W = 
(3\pi)^{1/3} \Big(\frac{3}{4}N\Big)^{4/3} \frac{1}{R},
\end{equation}
where
\begin{equation}\label{eqenergyden}
	W = \frac{\rm energy}{\rm  unit.volume} = \frac{2}{8\pi^3} \int_{0}^{k_{\rm F}} 4\pi k^2 \d k \cdot k = \frac{\omega_{\rm F}^4}{4\pi^2}.
\end{equation}
The energy associated with the scalar field $\phi$ is instead
\begin{equation}\label{eq:energyscalarfieldFS}
	E_s = \int_{0}^{R} 4\pi \rho^2 d\rho \hspace{0.1cm} (U + V) \simeq \Big(\frac{1}{6} \mu \phi_0^2\Big) 4\pi R^2\,,
\end{equation}
where we have used the fact that
\begin{equation}
    \frac{12}{\mu \phi_0^2} U \simeq  \frac{12}{\mu \phi_0^2} V \simeq  \delta(\rho - R)\,,
\end{equation}
which can be shown using Eq.~\eqref{eq:scalar_field_profile} and $\mu R\gg1$.

The total energy of our configuration is
\begin{equation}
	E = E_f + E_s,
\end{equation}
while the radius can be found by imposing ${\partial E}/{\partial R}=0$, yielding
\begin{equation}\label{eq:radius}
	R = \Big[\frac{3}{4\pi} (3\pi)^{1/3} \Big(\frac{3}{4}N\Big)^{4/3}\Big]^{1/3} \Big(\frac{1}{\mu \phi_0^2}\Big)^{1/3} 
\end{equation}
and the mass
\begin{equation}\label{eq:mass}
	M = E(R) = 12 \pi R^2  \Big(\frac{1}{6} \mu \phi_0^2\Big).
\end{equation}
From Eqs.~\eqref{eq:radius} and \eqref{eq:mass}, we get
\begin{equation}
	R \sim N^{4/9} \qquad M \sim N^{8/9}\,.
\end{equation}
Thus, at least for large $N$, the mass of the soliton is lower than the energy of the sum of $N$ free particles, ensuring stability.\footnote{This conclusion remains true also in the fully relativistic theory.}

In the absence of gravity, $M$ can be arbitrarily large. However, due to relativistic effects we expect the existence of a maximum mass beyond which the object is unstable against radial perturbations. 
We expect that gravity becomes important when $2 G M/R \sim 1$. Therefore, the critical mass $M_c$ can be estimated by simply imposing $R \sim 2 G M_c$ in Eq.~\eqref{eq:mass}, yielding $G^2 M_c \sim {1}/{\mu \phi_0^2}$ and thus
\begin{equation}\label{eq:mass_scaling}
    	 \frac{\mu M_c}{m_p^2} \sim \frac{1}{\Lambda^2}.
\end{equation}
Likewise, one can obtain the scaling of all other relevant quantities, 
which we collect in Table~\ref{eq:scalinglist}.
{
\renewcommand{\arraystretch}{1.6}
\setlength{\tabcolsep}{4pt}
\begin{table}[t]
\caption{Analytical scalings of some physical quantities at the maximum mass $M_c$ in the $\mu R\gg1$ limit.  }
\begin{tabularx}{1 \columnwidth}{|X|l|}
\hline
\hline
Mass
& 
${\mu M_c}/{m_p^2} \sim {1}/{\Lambda^2} $
\\
\hline
Radius
& 
$\mu R_c \sim  {\mu M_c}/{m_p^2}  \sim {1}/{\Lambda^2}$ 
\\
%
\hline
$\tilde{\omega}_{\rm F}$
& 
$\tilde{\omega}_{\rm F}^c
\sim  
(\mu/m_p)^{1/2}/(\phi_0/m_f) 
\sim \Lambda^{1/2}/{\eta} $ 
\\
\hline
Central pressure
& $\tilde{P}_c \sim  \tilde{\omega}_{\rm F}^4 \sim  {\Lambda^2}/{\eta^4}$
\\
\hline
\hline
\end{tabularx}
\label{eq:scalinglist}
\end{table}
}

\subsubsection{Self-consistency criteria}
When deducing the scaling reported in Table~\ref{eq:scalinglist}, 
 we  made the following assumptions:
\begin{itemize}
\item[{\it i)}] $\mu R \gg 1$;
\item[{\it ii)}] a gas of \emph{massless} fermions in the interior of the star.
\end{itemize}
In practice, the first assumption is not restrictive (see e.g. \cite{Freivogel:2019mtr}).
Indeed, since $\mu^{-1}$
is the Compton wavelength of the scalar boson, in the context of a \emph{classical} field theory we should always impose $\mu R \gg 1$. In other words, if $\mu R \simeq 1$ the quantum effects of the scalar field become important on the scale of the star and one cannot trust the classical theory anymore. 
The hypothesis $\mu R \gg 1$  is an essential ingredient in order to approximate the scalar field profile with Eq.~\eqref{eq:scalar_field_profile}, 
and to assume, as a consequence, 
that $k_{\rm F}$ is a step function. 
Besides, it guarantees that the energy density of the scalar field is near a delta function. 
Using the scaling reported in Table~\ref{eq:scalinglist},
condition~{\it i)} implies $ \Lambda \ll 1$.

One may worry that the second assumption can be violated,
since the scalar field is not located \textit{exactly at} $\phi_0$ 
in the origin $\rho = 0$, and therefore fermions are never exactly  massless. 
It is enough checking that the fermion gas is \textit{very close} to 
be a massless gas. 
Let us recall that the effective mass of the fermion is defined as
\begin{equation}
	m_\eff(\rho) = m_f\Big(1 - \frac{\phi(\rho)}{\phi_0}\Big) 
\end{equation}
and therefore $m_\eff(\rho = 0) = m_f \epsilon$. 
We can say that the fermion gas is effectively massless when $W/P = 3$. 
From Eqs.~\eqref{fermion_energy} and~\eqref{fermion_pressure}, at the lowest order in $\epsilon$ one obtains
\begin{equation}
	\dfrac{W}{P} = 3\Big(1 + \frac{2 m_f^2\epsilon^2}{k_{\rm F}^2}\Big) + O(\epsilon^3),
\end{equation}
which indicates we should require
\begin{equation}
	\frac{2 m_f^2\epsilon^2}{k_{\rm F}^2} \ll 1
\end{equation}
in the vicinity of the origin at $\rho \simeq 0$.
At larger radii, the scalar field gradually moves away from the central configurations and fermions start retaining a bare mass.
Inserting Eq.~\eqref{eq:kFermiFlatSpace} in the previous condition and expanding
Eq.~\eqref{Pcvsomegafepsilon} 
provide the condition we need to enforce to obey assumption {(\it ii)}, i.e.,
\begin{equation}\label{eq:secondfunhyp}
	\frac{2 m_f^2\epsilon^2}{(12\pi^2 P_c)^{1/2}} \ll 1.
\end{equation}
We express $\epsilon$ using the scalar field profile approximation in Eq.~\eqref{eq:scalar_field_profile}. Indeed, with simple manipulations, one finds
\begin{equation}\label{eq:tosubstitute}
	-\log \epsilon = \mu R \,  	\gg 1.
\end{equation}
Substituting~\eqref{eq:tosubstitute} in~\eqref{eq:secondfunhyp},
and neglecting, at this stage, the numerical factors one obtains
\begin{equation}
	   \log \Bigg(\frac{m_f}{P_c^{1/4}}\Bigg) \ll  \mu R.
\end{equation}
Using the scaling relations in Table~\ref{eq:scalinglist}, we obtain
\begin{equation}
	\log \Bigg(\frac{\eta}{\Lambda^{1/2}}\Bigg)\ll \frac{1}{\Lambda^2}.
\end{equation}

Summing up, the following conditions on the parameters
\begin{align}\label{eq:criterion}
\Lambda &\ll 1,  \\
\log \Bigg(\frac{\eta}{\Lambda^{1/2}}\Bigg) &\ll \frac{1}{\Lambda^2}\label{eq:criterion2}
\end{align}
are our self-consistency criteria to check if we are in a regime in which the scaling reported in Table~\ref{eq:scalinglist} is expected to be valid. 
While it can be shown that the second condition implies the first, we prefer writing both for the sake of clarity. 
Notice that, for fixed $\Lambda\ll1$, 
one can violate~\eqref{eq:criterion2} for increasing values of $\eta$, but only logarithmically.

\subsubsection{Confining and deconfining regimes}\label{eta_critic_section} 

An important consequence of the scalings collected in Table~\ref{eq:scalinglist} is that the critical mass and radius are independent of $\eta$ at fixed $\Lambda$. We shall call the region of the parameters space where this happens the \textit{confining regime} of the solutions. Indeed, in this regime the size of the soliton is dictated by the parameters of the scalar field, i.e. $\mu$ and $\phi_0$, regardless of the value of the fermion mass $m_f$. Physically, we expect that this would be the case when there exists a hierarchy between the scalar and fermion parameters. Since this hierarchy is measured by $\eta$, we expect that the confining regime exists only when $\eta$ is larger than a critical value, $\eta_c$.

To better clarify this point, we consider again Eq.~\eqref{eq:kFermi} for the Fermi momentum,
\begin{equation}
 k_{\rm F}^2(\rho) = \omega_{\rm F}^2e^{-2u(\rho)} - m_f\Big(1 - \frac{\phi(\rho)}{\phi_0}\Big)^2\,.
\end{equation}
In the $m_f \to 0$ limit this quantity becomes positive definite and so the fermionic pressure cannot vanish at any finite radius. In other words, the radius of the star can be arbitrarily large, provided that $m_f$ is sufficiently small. This is nothing but the well-known fact that a star made of purely relativistic gas does not exist.

Hence, if we enter a regime where the fermion bare mass $m_f$ is so small that, even after the scalar field has moved away from the false vacuum (where the effective fermion mass is small by construction), the Fermi gas is still relativistic, then the radius of the star grows fast and a small variation in $m_f$ produces a big variation in the radius. We call this regime the \textit{deconfining regime} of the solution.

In terms of the dimensionless variables  defined above, the $m_f \to 0$ limit becomes
\begin{equation}
    \tilde{\omega}_{\rm F}\to \infty.
\end{equation}
Therefore, we expect that, for a given choice of $(\Lambda, \eta)$, the confining regime exists only if $\tilde{\omega}^c_{\rm F}$ is smaller than a certain value. Using the scaling for $\tilde{\omega}^c_{\rm F}$ in Table~\ref{eq:scalinglist}, this can be translated into the condition
\begin{equation}\label{condition_on_eta}
    \frac{\Lambda^{1/2}}{\eta} < C,
\end{equation}
where $C$ is a constant that has to be determined numerically.

At this point, it is natural to define $\eta_c$ as the value of $\eta$ in which Eq.~\eqref{condition_on_eta} is saturated. In this way, Eq.~\eqref{condition_on_eta} becomes
\begin{equation}\label{condition_on_eta2}
   \eta > \eta_c = C \Lambda^{1/2}.
\end{equation}

To summarize, when $\eta\gtrsim\eta_c$ (confining regime) the size of the soliton near the maximum mass is mostly determined by the properties of the scalar field, whereas it strongly depends on the fermion mass when $\eta\lesssim\eta_c$ (deconfining regime\footnote{Note that, deep in the deconfining regime (when $\eta\to0$), the Compton wavelength of the fermion, $1/m_f$, might become comparable to or higher than the radius of the star. In this case we expect the Thomas-Fermi approximation to break down.}).

\subsection{Energy conditions}
For an energy-momentum tensor of the form
\begin{equation}
	T^\mu_\nu = \text{diag}\{-\rho, P_1, P_2, P_3 \},
\end{equation}
the energy conditions take the following form:
\begin{itemize}[leftmargin=*]
    \item Weak energy condition:
    $\rho \geq 0 \text{ and } \rho + P_i \geq 0.$
    \item Strong energy condition:
    $\rho + \sum_{i} P_i \geq 0\text{ and } \rho + P_i \geq 0$.
    \item Dominant energy condition:
    $\rho \geq \abs{P_i}$.
\end{itemize}
For a spherically symmetric configuration, $P_1=P_r$ is the radial pressure, while $P_2=P_3=P_t$ is the tangential pressure. For our model,
\begin{align}
 \rho & = U+V+W\,, \\
 P_r & = V-U+P\,, \\
 P_t & = -U-V+P\,.
\end{align}
Since $V,W,P$ are nonnegative quantities, we obtain $\rho+P_r\geq0$ and $\rho+P_t\geq0$.
Thus, the weak and strong energy conditions are satisfied if
\begin{align}
	U+V+W &\geq 0 ,
	\\
	3P-2U+W &\geq 0 \,,
\end{align}
respectively.
Since $U$ is also a non-negative quantity, the weak energy condition is always satisfied, while the strong energy condition can be violated. In particular, it is violated even in the absence of fermions ($P=W = 0$).

\begin{figure*}[!t] 
    \centering
    \includegraphics[width=0.49\linewidth]{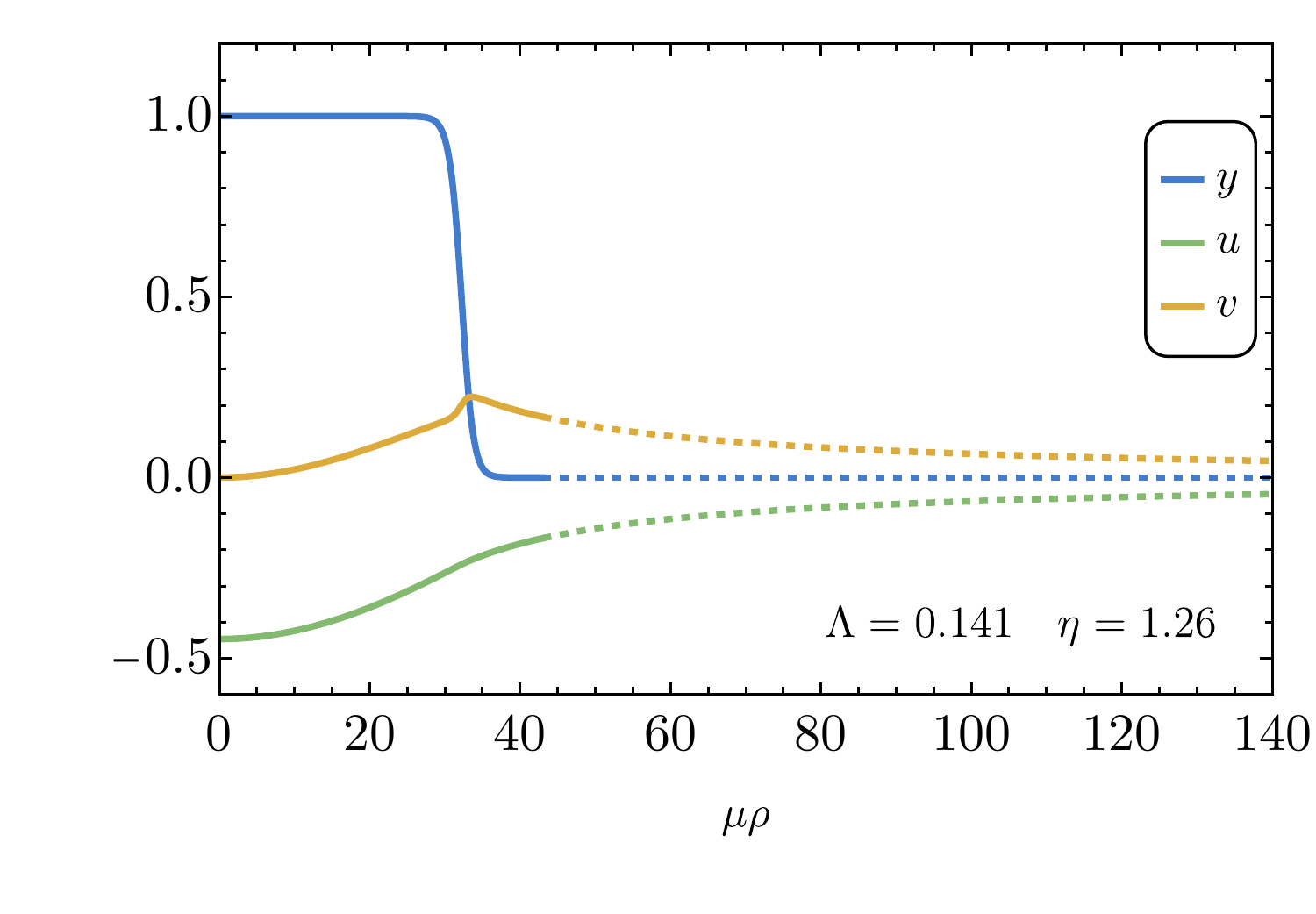}%
     \includegraphics[width=0.49\linewidth]{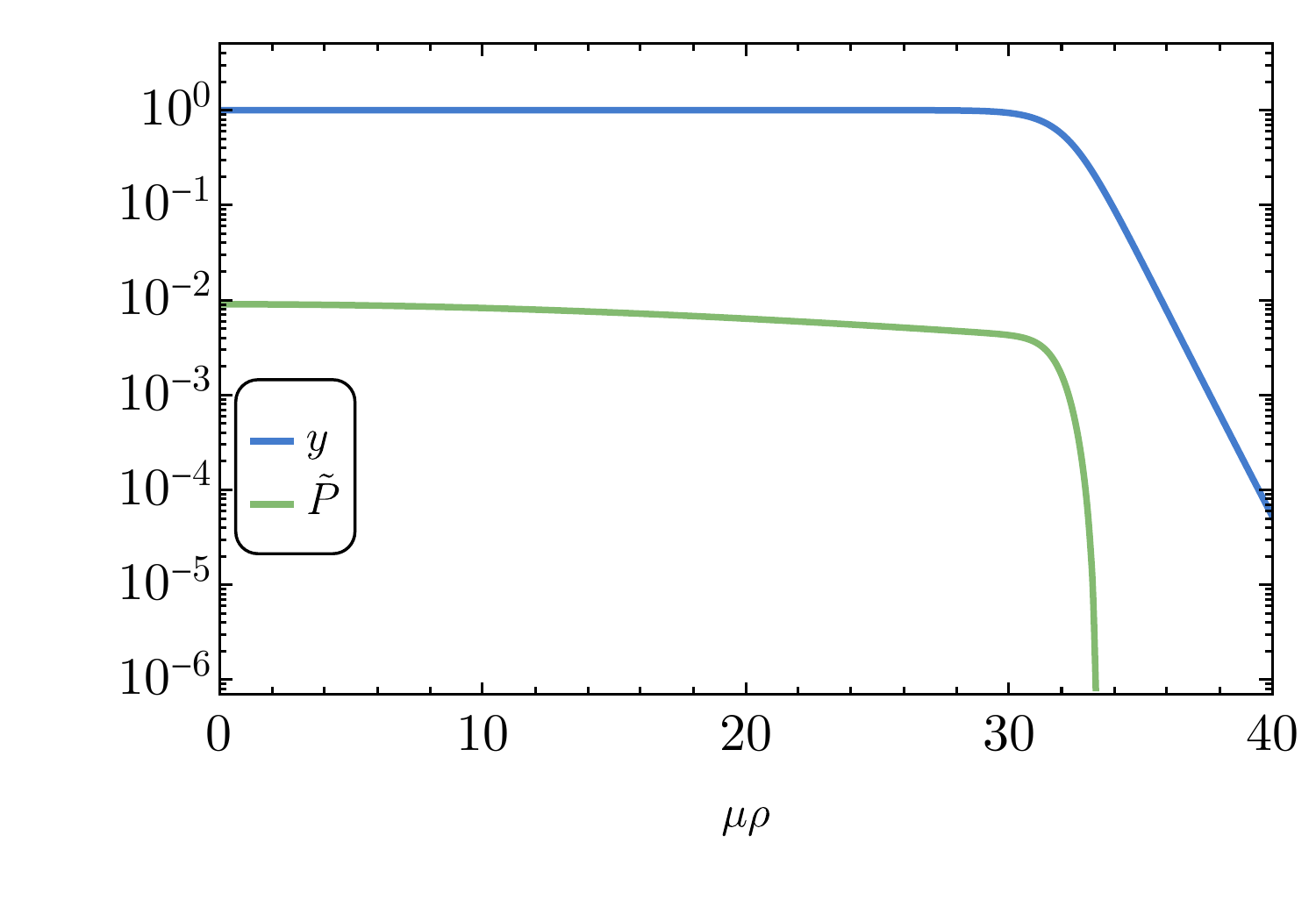}%
     
    \includegraphics[width=0.49\linewidth]{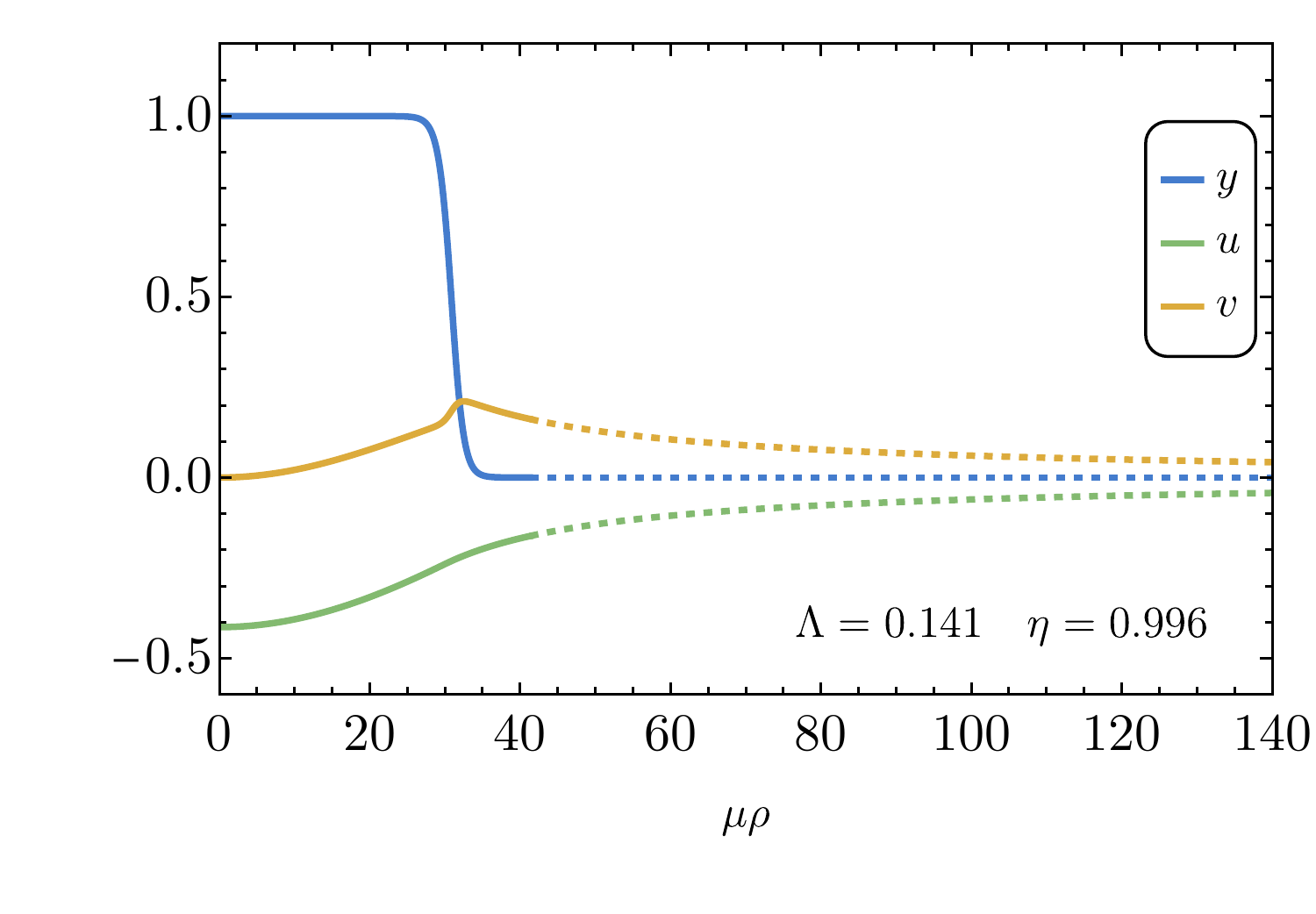}%
     \includegraphics[width=0.49\linewidth]{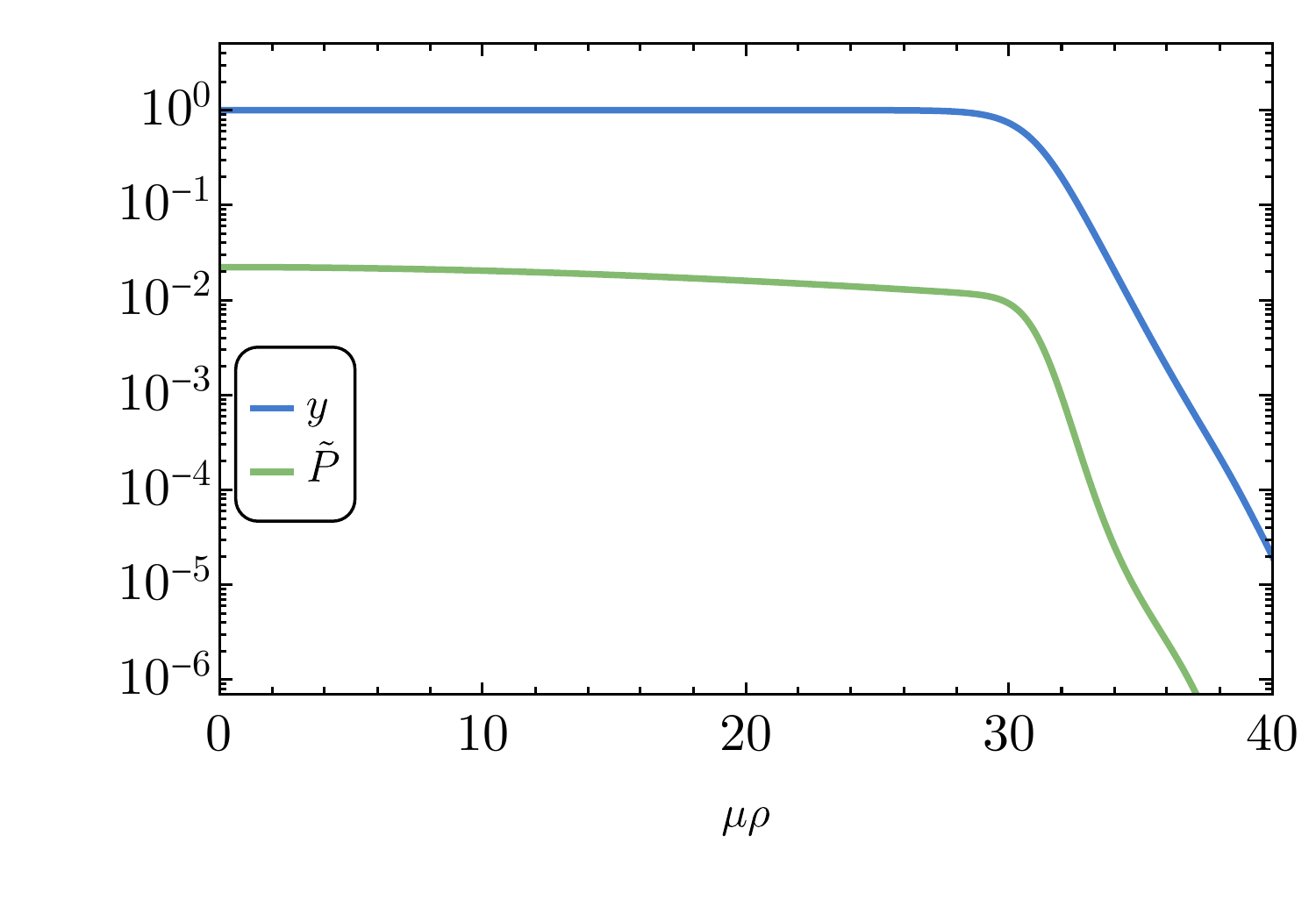}%
    \caption{
    Radial profiles of the adimensional pressure $\tilde P$, scalar profile $y$ and metric functions $u$ (shifted) and $v$
    for two example configurations. Continuous lines represent numerical data, whereas dashed lines reconstruct the asymptotic behavior of the solutions by fitting with the Schwarzschild solution.
    {\bf Top panels:}
     $\Lambda = 0.141$, $\eta = 1.26$, 
     $\tilde{P}_c = 0.00903$, and $ \log_{10}\epsilon = -13.9$. 
     The mass and radius of the soliton fermion star are $\mu M / m_p^2 =6.14$ and $\mu R=33.8$, respectively.
     This solution falls within the confining regime.
     {\bf Bottom panels:}
      $\Lambda = 0.141$, $\eta = 0.996$,
      $\tilde{P}_c = 0.0222$, and $ \log_{10}\epsilon = -12.9$.
      The mass and radius of the soliton fermion star are $\mu M / m_p^2 =5.71$ and 
    $\mu R=39.3$,
      respectively.
      This solution falls within the deconfining regime.
     }
    \label{fig:profiles}
\end{figure*}

The dominant energy condition, instead, gives two inequalities:
\begin{align}
	&U+V+W \geq \abs{P+V-U},
	\\
	&U+V+W \geq \abs{P-V-U}.
\end{align}
One can show that the dominant energy condition is satisfied  whenever 
\begin{equation}\label{key}
	W+2(U+V)\geq P,
\end{equation}
This inequality is satisfied if
\begin{equation}\label{key2}
	W - P \geq 0\,,
\end{equation}
which can be shown to be true using the analytic expressions of $W$ and $P$.

To sum up, the weak and dominant energy conditions are always satisfied, while the strong energy condition can be violated (e.g. in the absence of fermions) as generically is the case for a scalar field with a positive potential~\cite{Herdeiro:2019oqp}.

\begin{figure*}[t] 
    \centering
    \includegraphics[width=0.49\linewidth]{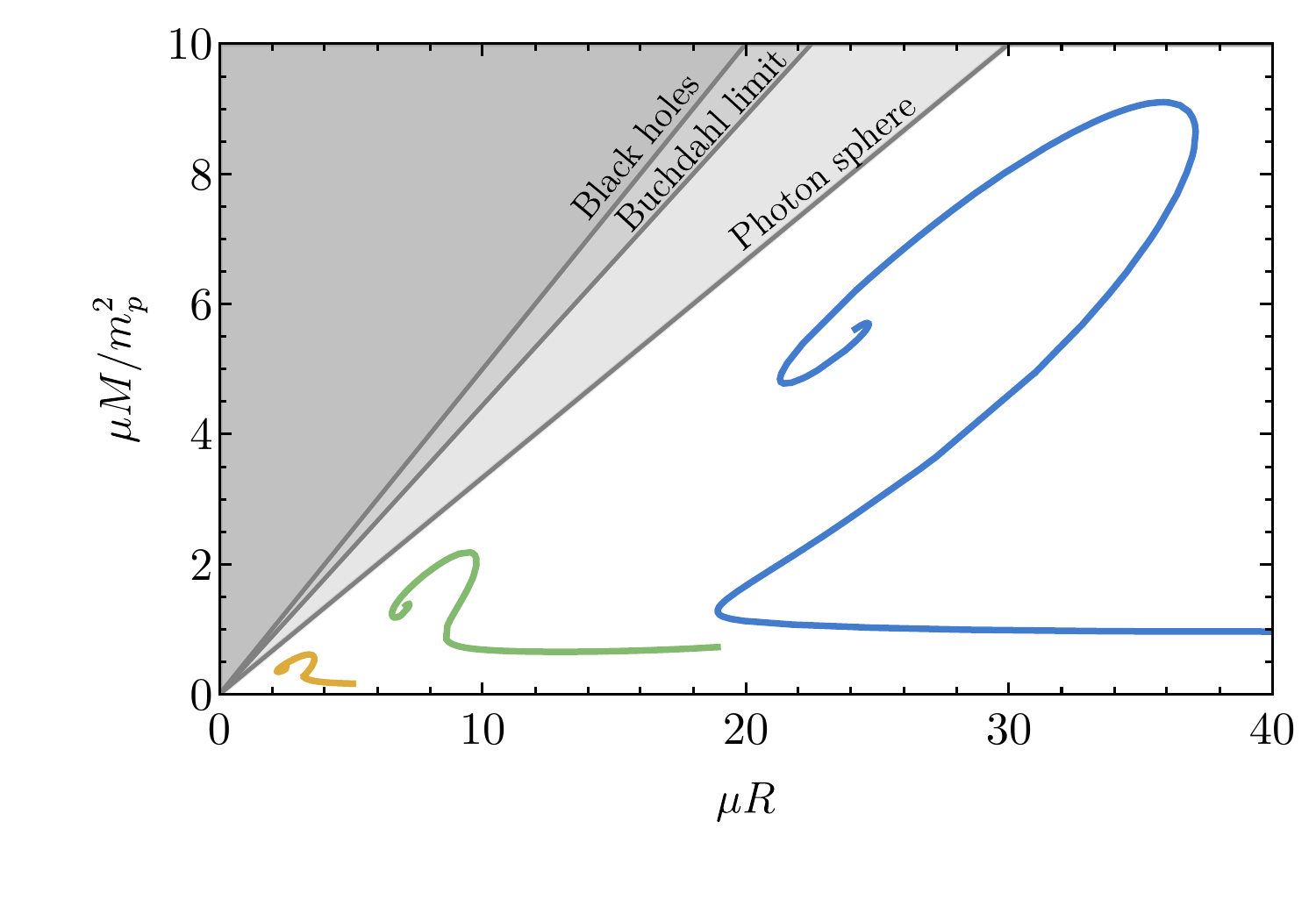}%
    \includegraphics[width=0.49\linewidth]{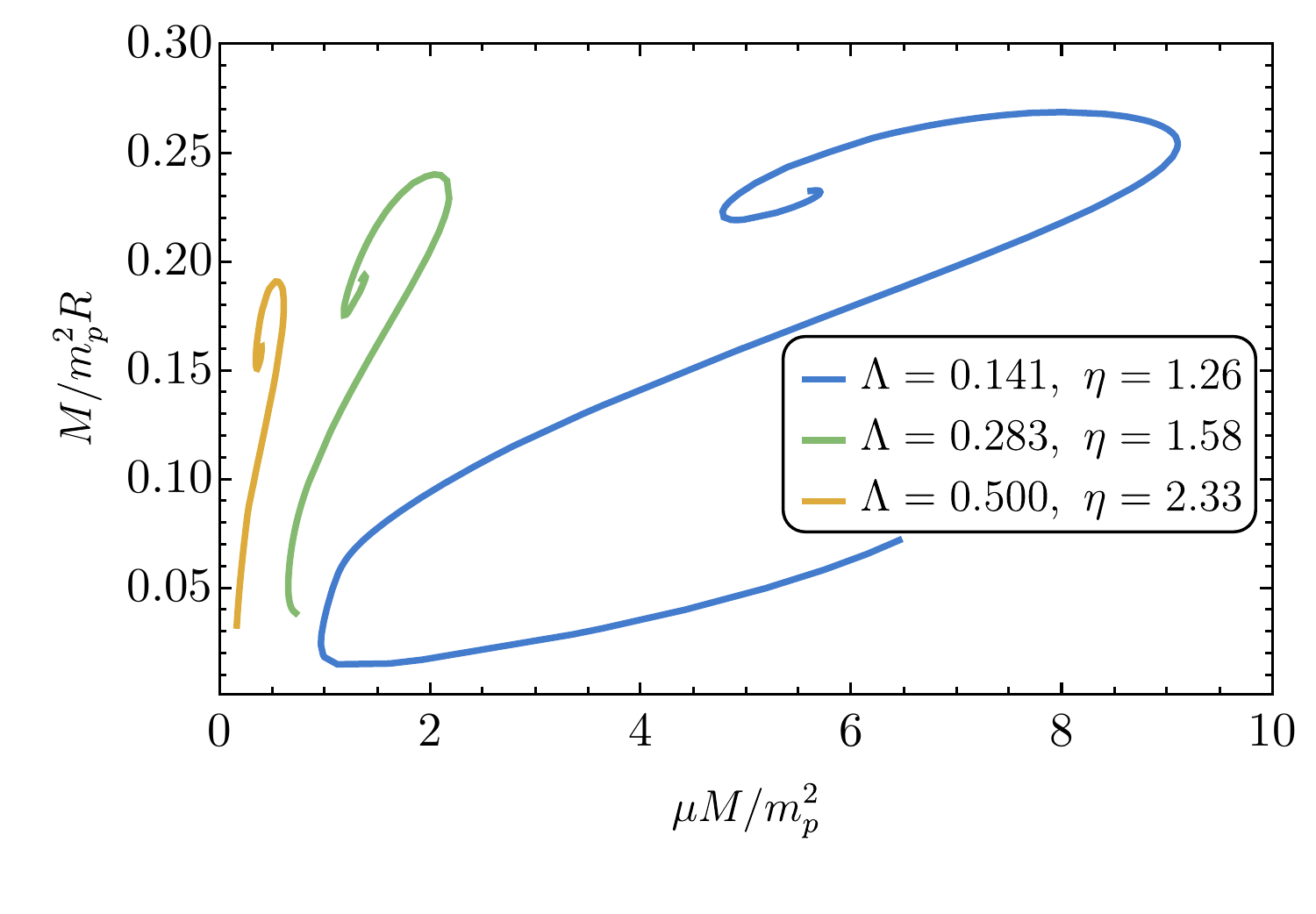}
    
    \includegraphics[width=0.49\linewidth]{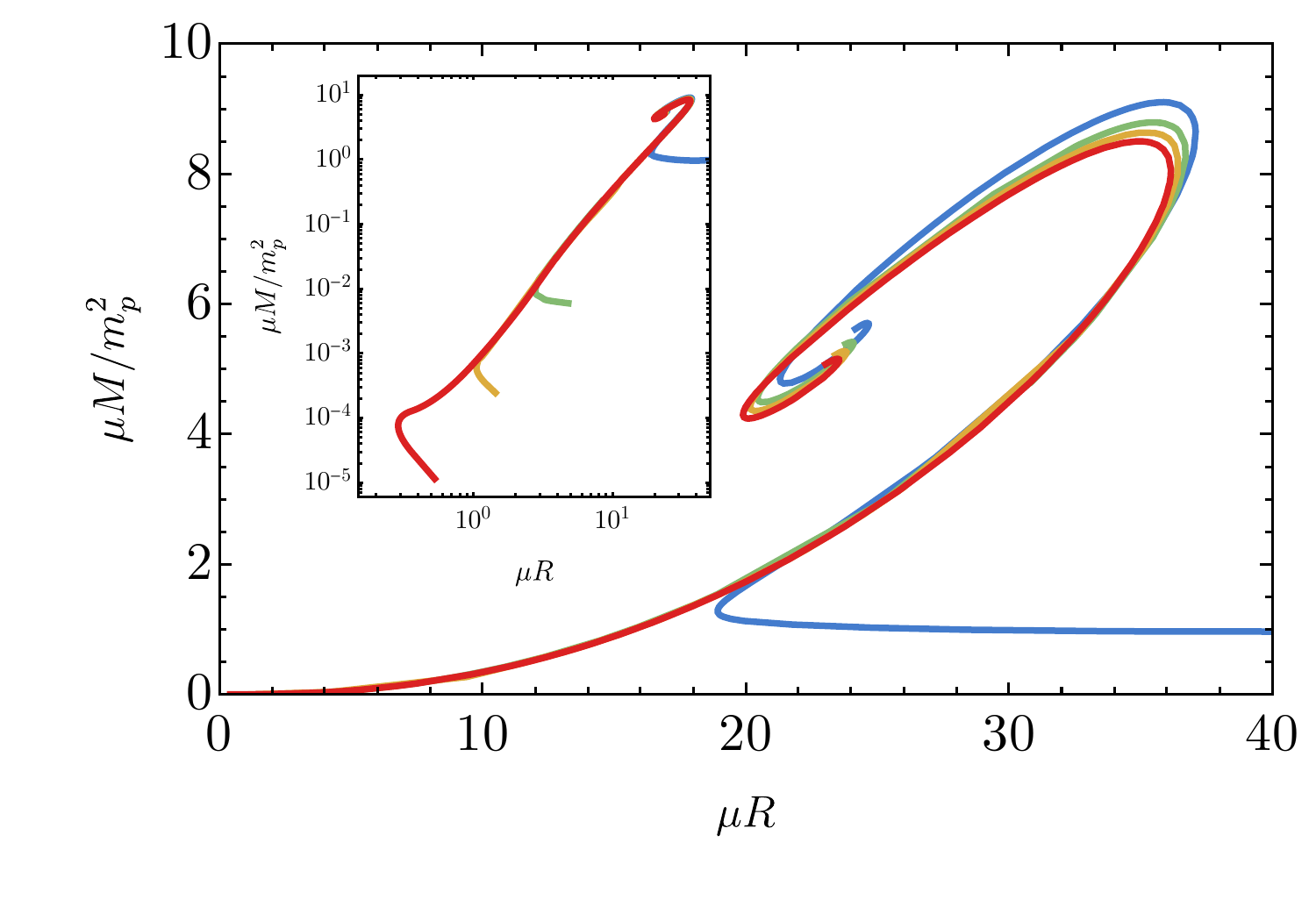}%
    \includegraphics[width=0.49\linewidth]{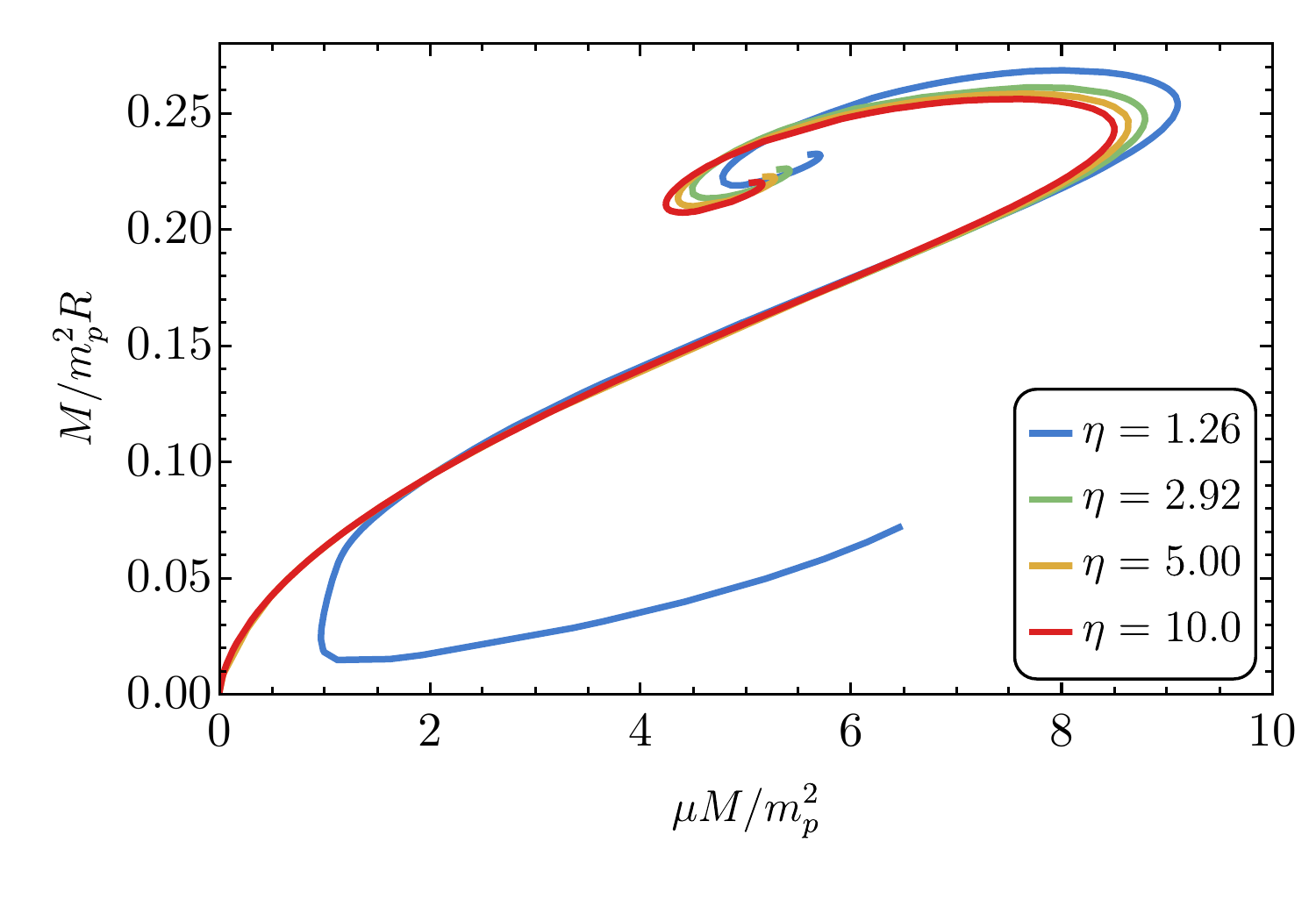}%
    \caption{Mass-radius (left panels) and compactness-mass (right panels) diagrams for fermion soliton stars.  
    The top panels refer to various values of $(\Lambda,\eta)$ in the confining regime ($\eta>\eta_c$; see Sec.~\ref{eta_critic_section}). 
    As a reference, in the top-left panel we also draw the lines $R=2\,GM$, $R=9/4\,GM$, $R=3\,GM$, corresponding to the Schwarzschild radius, Buchdhal's limit~\cite{Buchdahl:1959zz}, and the photon-sphere radius.
    The bottom panels refer to various values of $\eta$ for fixed $\Lambda = 0.141$. The smallest value of $\eta$ considered is near but greater than the critical value.
    The inset shows the curves in logarithmic scale, to highlight that in this case there exists a turning point in the $M$-$R$ diagram at low masses
    that proceeds toward the Newtonian limit of small $M$ and large $R$.
    }
    \label{fig:massradious}
\end{figure*}

\section{Numerical results}\label{sec_num_res}

In this section, we present the fermion soliton solutions in spherical symmetry obtained by integrating the field equations~\eqref{fund_sistema_dimensionlesse_dimensionlesse}.
We will confirm the existence of a solution beyond the thin-wall approximation used in Ref.~\cite{Lee:1986tr}.
Also, based on the numerical solutions, we are able to confirm the scalings derived in the previous sections in a certain region of the parameter space and fix their prefactors.

\subsection{Numerical strategy}

In this section, we summarize the numerical strategy we adopt to find soliton fermion solutions.
Given the boundary condition 
\eqref{initial_conditions_dimensionlessi}, 
the set of equations 
\eqref{fund_sistema_dimensionlesse_dimensionlesse} are solved numerically by adopting the following strategy:
\begin{enumerate}
\item We fix a certain value of $\tilde\omega_{\rm F}$;
\item for a given value of $\tilde\omega_{\rm F}$ and of the central scalar field (i.e., a value of $\epsilon$), we obtain $\tilde P_c$ through Eq.~\eqref{Pcvsomegafepsilon},
and therefore $x$ through the last equation in~\eqref{fund_sistema_dimensionlesse_dimensionlesse}\footnote{Equivalently, one can give initially $\tilde{P}_c$, $\epsilon$ and determine $\tilde{\omega}_{\rm F}$ inverting Eq.~\eqref{Pcvsomegafepsilon}.};
\item 
we integrate the first three equations in~\eqref{fund_sistema_dimensionlesse_dimensionlesse} for the variables $(u, v, y)$, starting from $r \approx 0$ to the point $r=R_f$ where the fermion pressure drops to negligible values,  $\tilde{P}(R_f) = 0$;
\item we eliminate the fermionic quantities from the system of equations~\eqref{fund_sistema_dimensionlesse_dimensionlesse} and 
start a new integration with initial conditions given at $r = R_f$ imposing continuity of the physical quantities. That is, the initial conditions on the metric and scalar fields at $r = R_f$ are obtained from the last point of the previous integration up to $r=R_f$;
\item we use a shooting method 
to find the value of $\epsilon$ 
that allows an asymptotically flat solution to exist, which means imposing $y(r \to \infty) \to 0$;
\item as previously discussed, because the scalar field does not have a compact support, we define the radius of the star ($R>R_f$) as that containing $99\%$ of the total mass, i.e. $\tilde{m}(R)=0.99 \,\mu M / m_p^2$ (Eq.~\eqref{eq:defmass}), and the compactness is $GM/R$;
\item Finally, we repeat the procedure for a range of values of $\tilde\omega_{\rm F}$, finding a one-parameter family of solutions. As we shall discuss, in certain regimes (including the deconfining one) this family exists only if $\tilde P_c$ is above a certain threshold, therefore lacking a Newtonian limit.
\end{enumerate}

As already noted, a vanishing scalar field $(y = 0, \partial_r y = 0)$ is a solution to the scalar equation in Eq.~\eqref{fund_sistema_dimensionlesse_dimensionlesse} only if $S = 0$, that is, in the absence of fermions. This ensures that in any solution with $y\to0$ at infinity the fermion pressure must vanish at some finite radius.
Therefore, the fermion soliton solution is described by a fermion fluid confined at $r\leq R_f$ and endowed with a real scalar field that is exponentially suppressed outside the star, as expected from the discussion in Sec.~\ref{section_someptc}.

As described in the previous section, important parameters are the mass and radius of the critical solutions, $M_c$ and $R_c$. In practice, we compute these quantities by identifying in the $M$-$R$ diagram the point of maximum mass.

\begin{figure*}[t] 
    \centering
    \includegraphics[width=0.49\linewidth]{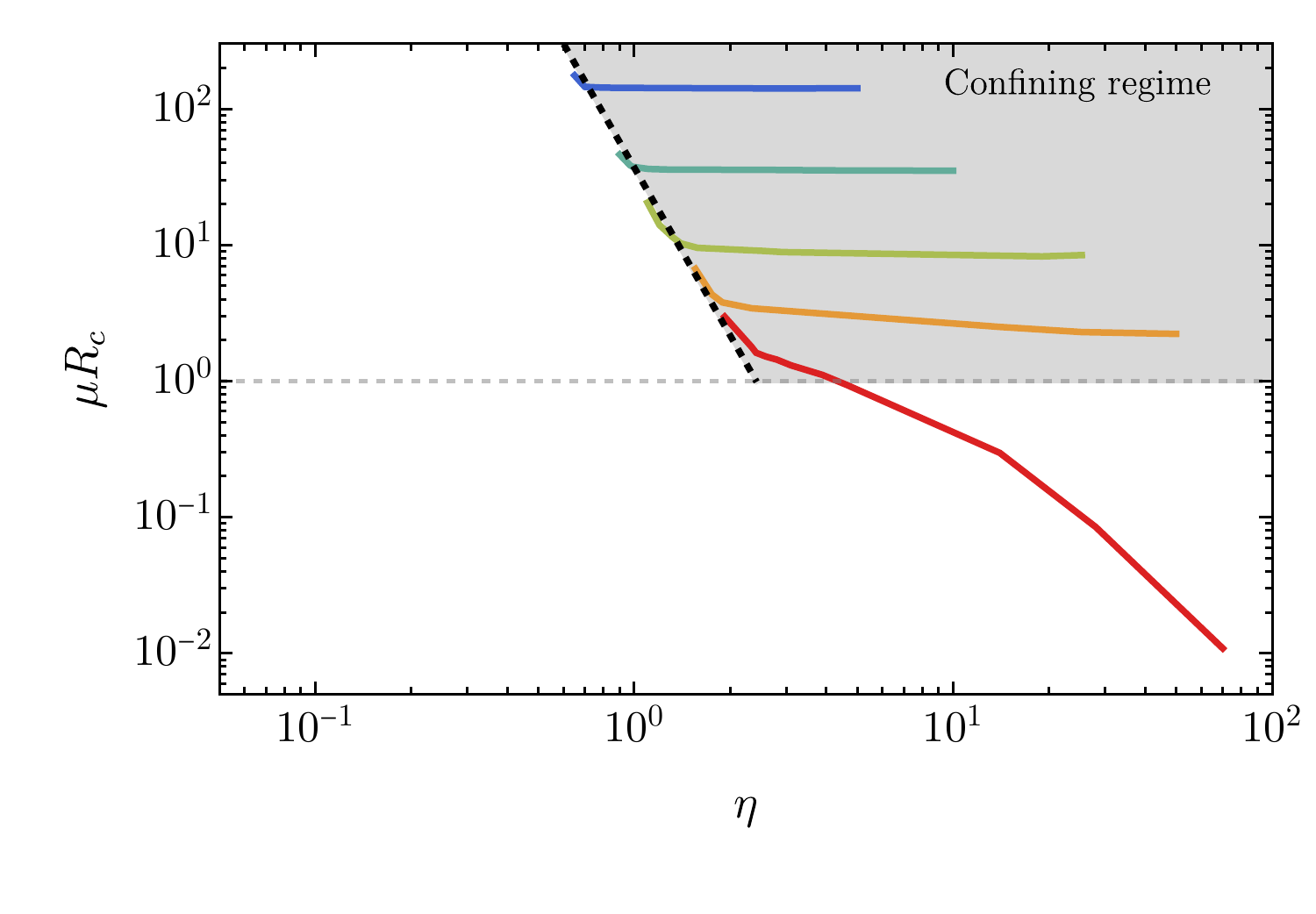}%
    \includegraphics[width=0.49\linewidth]{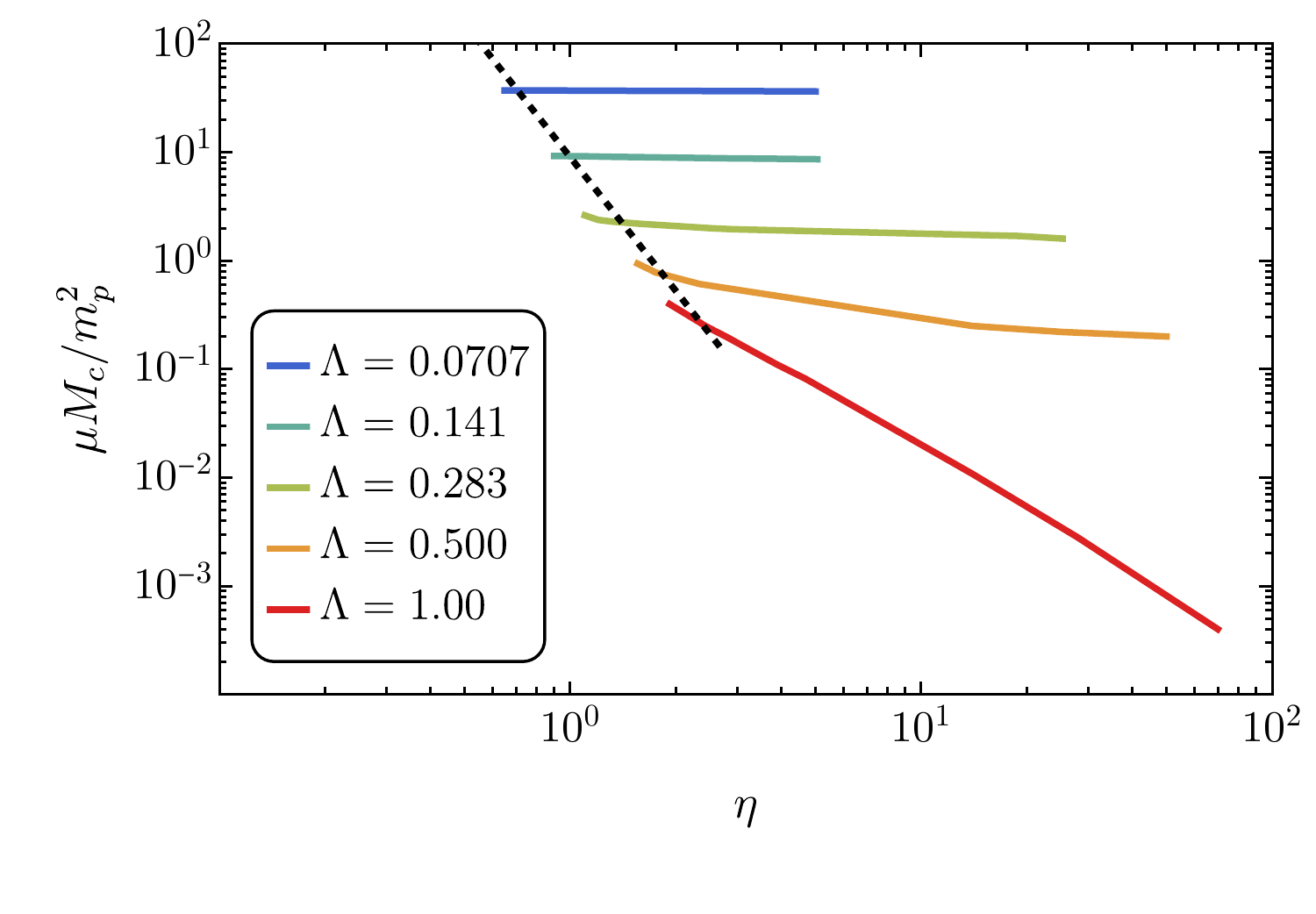}
    \caption{\textbf{Left:} Behavior of the critical radius $R_c$ with $\Lambda$ and $\eta$. The scaling~\eqref{condition_on_eta2} is highlighted by the diagonal black dashed line. 
    We observe an agreement until $\Lambda \lesssim 0.3$ whereas, for larger $\Lambda$,
    $\eta_c$ increasingly exceeds the predicted value. The horizontal grid line highlights 
    when the $\mu R > 1$ regime ends. 
    The shaded region above the two dashed lines is the confining regime.  \textbf{Right:} Behavior of the critical radius $M_c$ with $\Lambda$ and $\eta$. We observe that the critical mass does not exhibit a significant change of behavior for $\eta < \eta_c$.
    }\label{fig:scaling1}
\end{figure*}

\subsection{Fermion soliton stars} 
First of all, we confirm that fermion soliton stars exist also beyond the thin-wall approximation used in Ref.~\cite{Lee:1986tr}. An example is shown in Fig.~\ref{fig:profiles} which presents the radial profiles for the metric, scalar field, and fermion pressure.

Inspecting the panels of Fig.~\ref{fig:profiles} can help us understand
the qualitative difference between solutions in the confining regime (top) and the deconfining one (bottom). 
In the first case, as soon as the scalar field moves away from 
its central value at $\rho \to 0$,
and the effective mass of the fermion field grows,
the pressure quickly drops to zero. 
This reflects in the fact that the macroscopic size of the star $R$
is found to be very close to where the scalar field starts moving away from the false vacuum. This is the reason why the macroscopic properties of the 
star are mainly dictated by the scalar field potential. 
In the latter case, the small bare mass of fermions
makes them remain ultra-relativistic even when the 
scalar field moves away from the false vacuum, 
generating a layer where fermionic pressure drops exponentially but remains finite.
After the energy of fermions has fallen within the non-relativistic regime, 
fermionic pressure rapidly vanishes.
The existence of such a layer makes the final mass and radius of the star 
dependent on the fermion mass, see more details below. 
Also, as the numerical shooting procedure requires 
matching the asymptotic behavior of the scalar field 
outside the region where the energy density of the fermions remains sizable, 
deconfining solutions are characterized by a larger tuning
of the parameter controlling the central displacement $\epsilon$.

In Fig.~\ref{fig:massradious} we present the mass-radius and compactness-mass diagrams for various values of $\Lambda$ and $\eta$, in the confining regime. In the top panels, we observe that $\Lambda$ strongly affects the mass-radius scale and the maximum mass, while from the bottom panels we observe that $\eta$ has a weaker impact on the maximum mass, as expected from the discussion in Sec.~\ref{section_someptc}.

The dependence of $M_c$ and $R_c$ on $\Lambda$ and $\eta$ is presented in Fig.~\ref{fig:scaling1}. As expected, we observe that, for a fixed $\Lambda$, there is a critical value of $\eta$, below which the radius begins to grow rapidly. For $\eta > \eta_c$ and $\Lambda \lesssim 0.5$, we observe that the predictions given in Sec.~\ref{section_someptc} are valid, confirming the existence of a confining regime. Indeed, in that region of the parameter space, both the mass and the radius have a little dependence on $\eta$. This dependence grows very slowly for an increasing value of $\eta$, in agreement with Eq.~\eqref{eq:criterion2}. Moreover, the value of $\eta_c$ scales, for $\Lambda \lesssim 0.3$, in agreement with Eq.~\eqref{condition_on_eta2}, while for larger values of $\Lambda$ it exceeds the analytical scaling. At variance with the critical radius, the critical mass does not exhibit a change of behavior for $\eta < \eta_c$. As a consequence, the compactness decreases quickly.

In general, taking into account all the configurations numerically found, $\log_{10} \epsilon$ lies in the interval $(-150, -0.01)$.

Finally, in Table~\ref{measured_coefficients} we report the scaling coefficients computed numerically, which are valid in the confining regime ($\eta \gtrsim \eta_c$, $\Lambda \lesssim 0.5$).

\subsection{On the existence of a Newtonian regime}

From the bottom panels of Fig.~\ref{fig:massradious}, we observe that, even though $\eta$ has a weak impact on the maximum mass, it can qualitatively change
the $M-R$ diagram, especially at low masses. 
Overall, the mass-radius diagram reassembles that of solitonic boson stars~\cite{Friedberg:1986tq,Palenzuela:2017kcg,Bezares:2022obu,Boskovic:2021nfs} with several turning points in both the mass and the radius, giving rise to multiple branches (see also~\cite{Guerra:2019srj}).
The main branch is the one with $M'(R)>0$ before the maximum mass, which is qualitatively similar to that of strange (quark) stars~\cite{Alcock:1986hz,Urbano:2018nrs}.
However, the low-mass behavior (and the existence of a Newtonian regime) depends strongly on $\eta$.

\begin{figure*}[t] 
    \centering
    \includegraphics[width=0.46\linewidth]{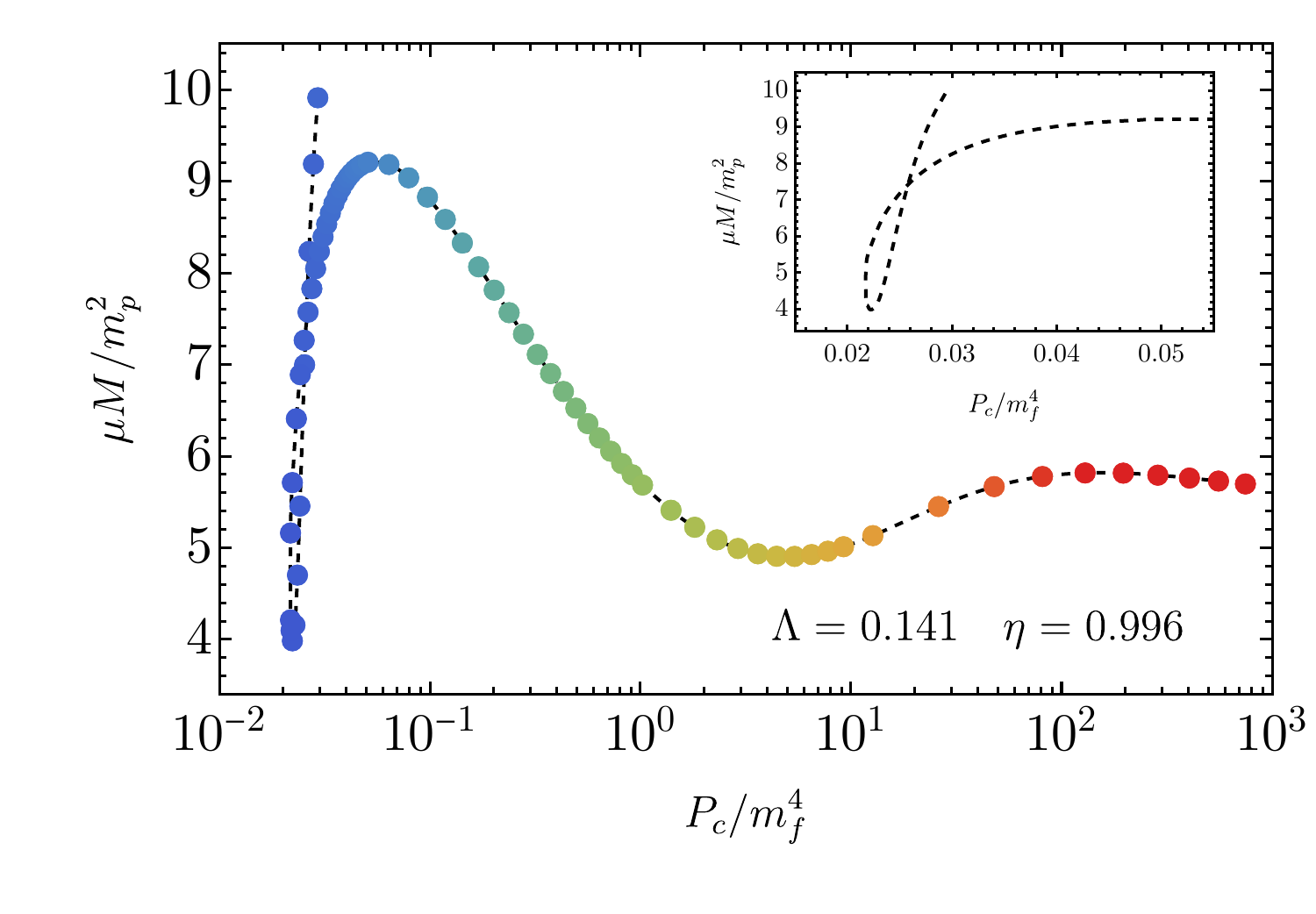}%
     \includegraphics[width=0.46\linewidth]{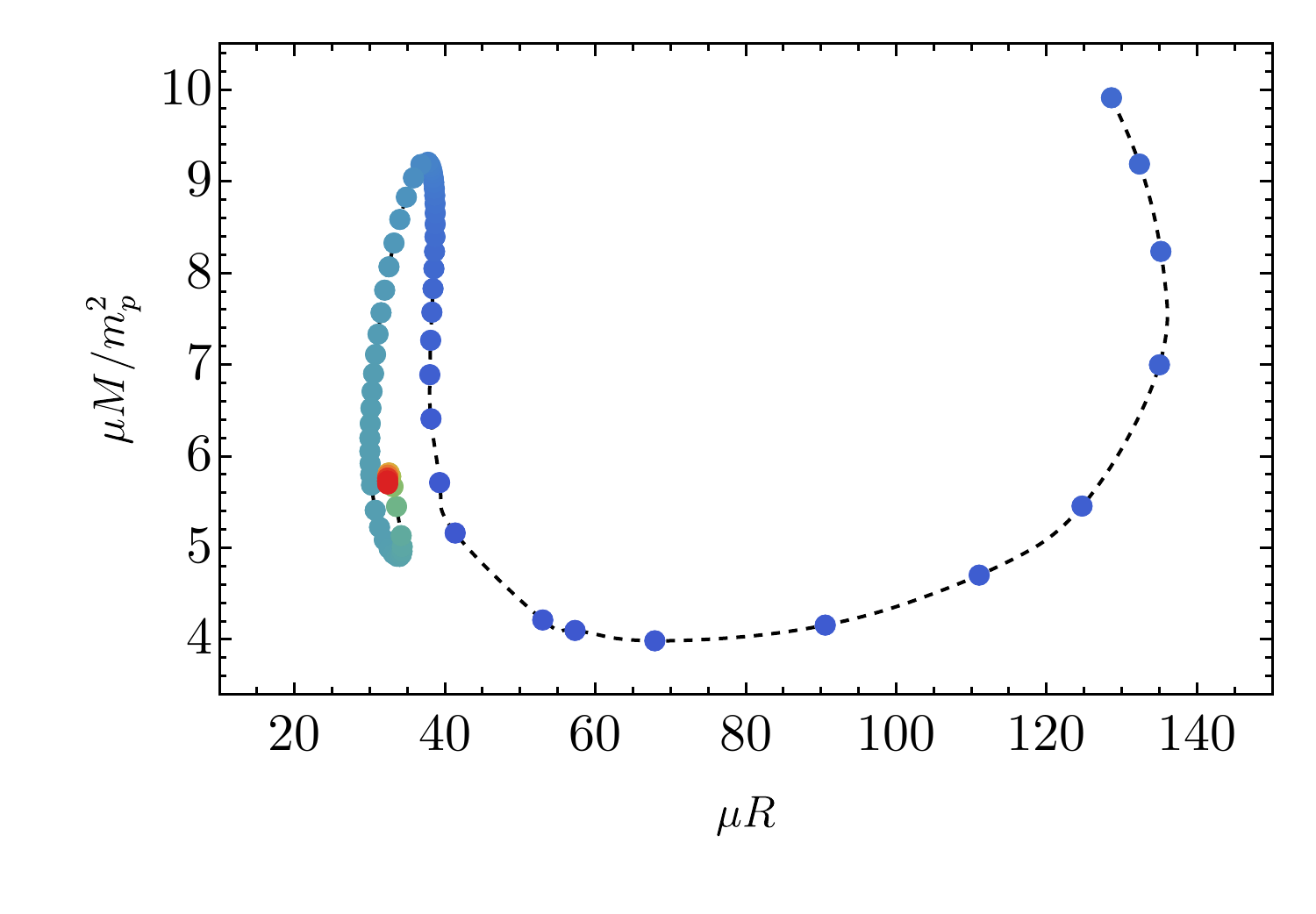}
     
     \includegraphics[width=0.46\linewidth]{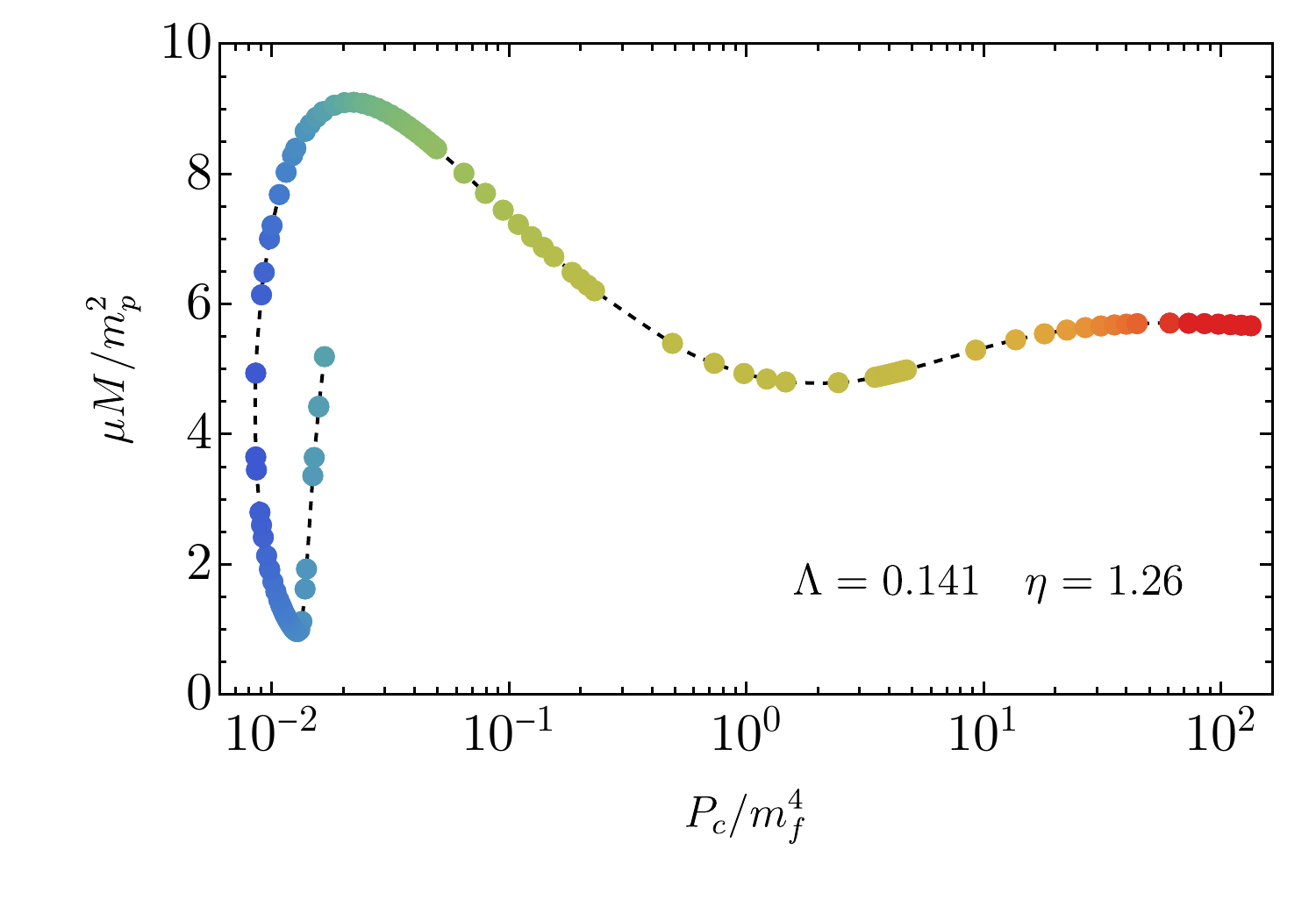}%
     \includegraphics[width=0.46\linewidth]{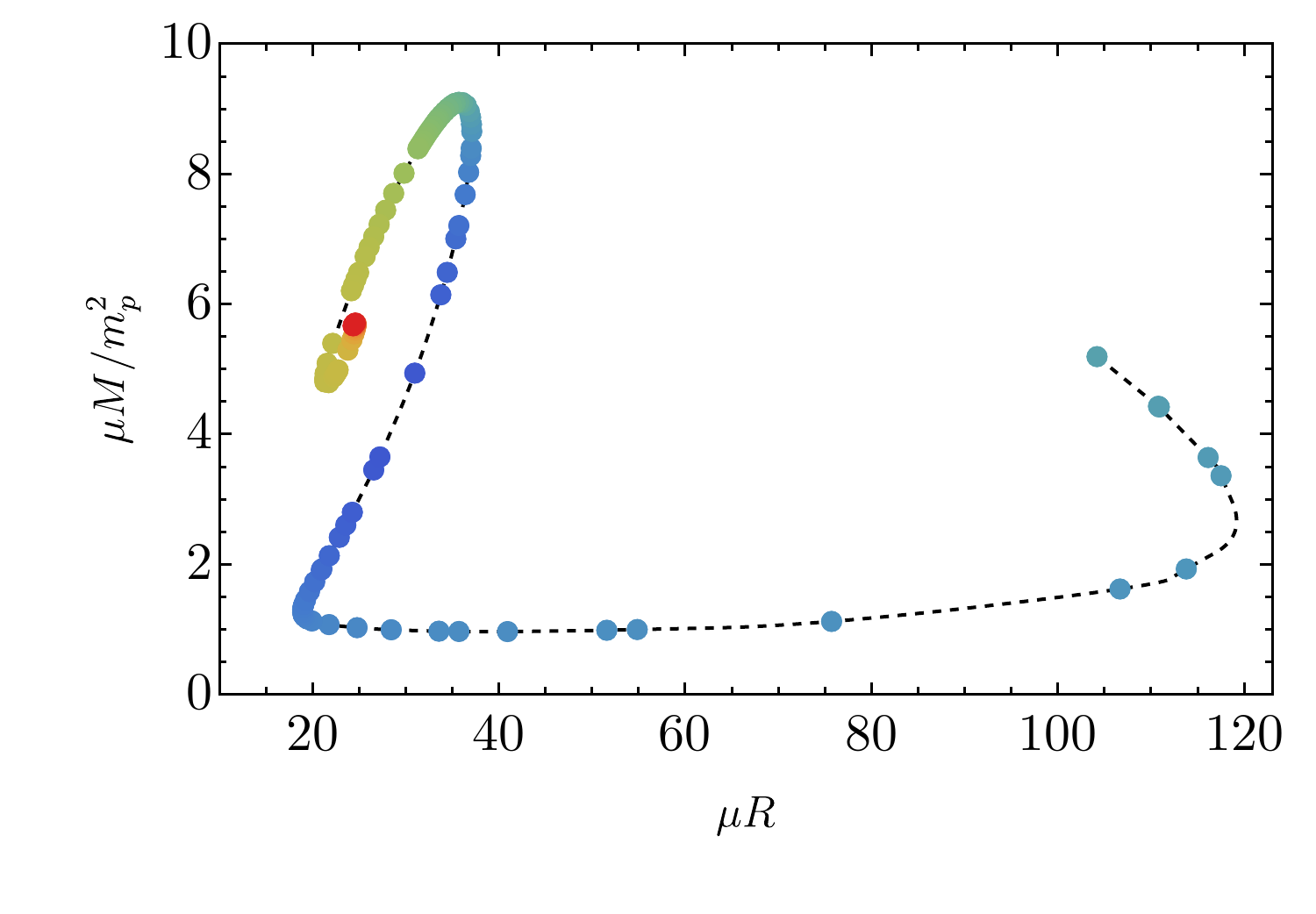}%
     
     \includegraphics[width=0.46\linewidth]{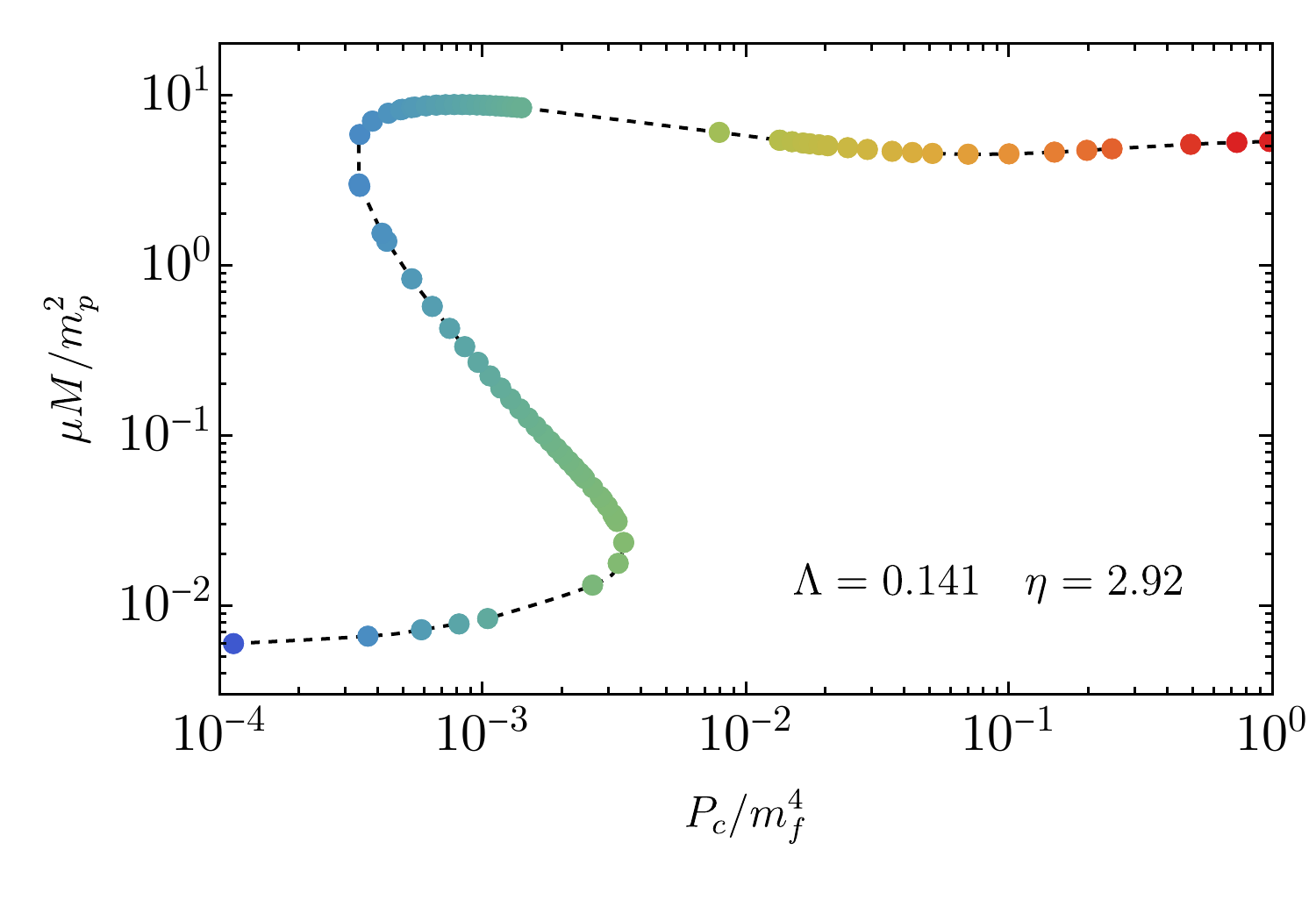}%
     \includegraphics[width=0.46\linewidth]{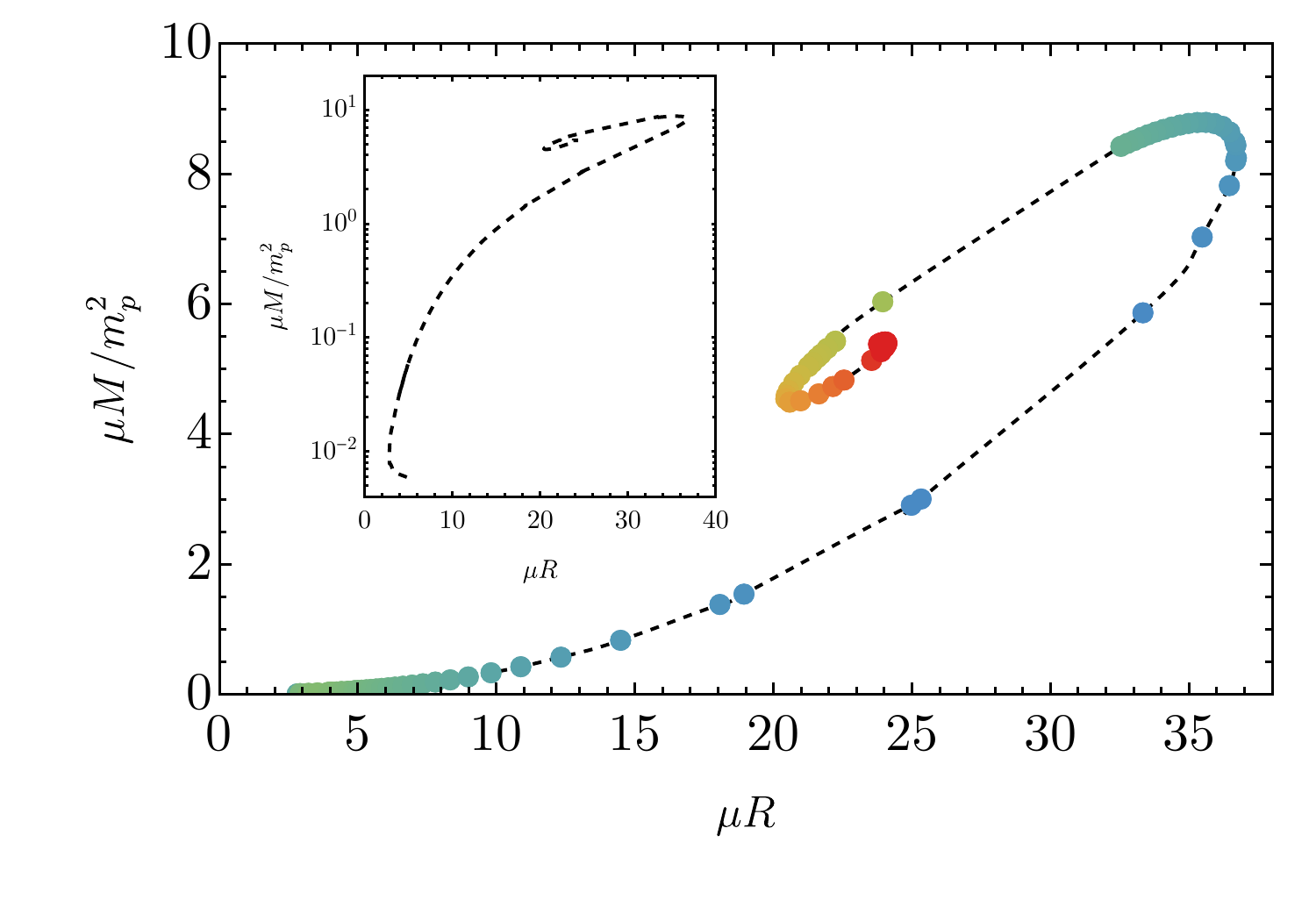}%
\caption{
\textbf{Left panels:} The mass of fermion soliton stars as a function of the central fermionic pressure. 
\textbf{Right panels:} The corresponding mass-radius diagram using the same color scheme as in the left panels, in order to associate with each point the corresponding central pressure.
\textbf{Top:} $\Lambda = 0.141$ and $\eta = 0.996$. This solution is in the deconfining regime and there is a lower bound on $\tilde{P}_c$ below which no solution exists. 
\textbf{Center:} $\Lambda = 0.141$ and $\eta = 1.26$. This solution is in the confining regime but, also in this case, there exists a lower bound on $\tilde{P}_c$.
\textbf{Bottom:} $\Lambda = 0.141$ and $\eta = 2.92$. This solution is in the confining regime but, given the larger value of $\eta$, there is no lower bound on $\tilde{P}_c$ and a Newtonian regime exists.
In all three cases, for a certain range of $\tilde{P}_c$ there are multiple solutions with the same central fermionic pressure and different central value of the scalar field. 
} \label{fig:central_pressurevsmass}
\end{figure*}

For sufficiently large values of $\eta$ (always in the confining regime) there exists a low-compactness branch in which $M'(R)<0$ and where the fermionic pressure is small compared to the energy density, giving rise to a Newtonian regime. 
However, an interesting effect 
starts occurring for values of $\eta$ near, but greater than, the critical one (e.g., the blue curve for $\eta=1.26$ in the bottom panels of Fig.~\ref{fig:massradious}\footnote{Notice that, in the bottom left panel, it is not possible to see the complete tail of the $M$-$R$ diagram. As underlined in the text, in the center right panel of Fig.~\ref{fig:central_pressurevsmass} we plot the complete $M$-$R$ diagram.}) all the way down to the deconfining regime. In this case, there is still a lower turning point in the $M$-$R$ diagram, but the compactness eventually starts \emph{growing} (see right bottom panel).
In this case there is no Newtonian regime, since the compactness is never arbitrarily small.

{
\renewcommand{\arraystretch}{1.6}
\setlength{\tabcolsep}{4pt}
\begin{table}[t]
\caption{Various scaling of the 
critical parameters with coefficients derived numerically in the $\Lambda\lesssim 0.5$ range. }
\begin{tabularx}{0.98 \columnwidth}{|l|l|}
\hline
\hline
Critical mass
& 
${\mu M_c}/{m_p^2} \approx 0.19 / \Lambda^2 $
\\
\hline
Critical radius
& 
$\mu R_c\approx 0.71 / \Lambda^2 $ 
\\
%
\hline
Compactness of the critical solution
& 
$C_{c} \approx 0.27$
\\
\hline
Critical value of the scale ratio
& $\eta_{c} \approx 2.7 \, \Lambda^{1/2}$ 
\\
\hline
\hline
\end{tabularx}
\label{measured_coefficients}
\end{table}
}

This peculiar behavior is also related to another important feature of the model, namely the fact that, for $\eta$ sufficiently small, fermion soliton stars exist only above a minimum threshold for the central fermionic pressure. We clarify this point in Fig.~\ref{fig:central_pressurevsmass}
. 
In the left panels we show the mass of the star as a function of the central fermionic pressure for $\Lambda=0.141$ and three values of $\eta$. For $\eta=0.966$ and $\eta=1.26$ (top and center panels), the pressure has a lower bound, corresponding to the absence of a Newtonian limit. For $\eta = 2.92$ (bottom panels) the behavior is qualitatively different and in this case the Newtonian regime is approached as $P_c\to0$.

To clarify where the minimum pressure and these multiple branches are in the mass-radius diagram, in the right panels of Fig.~\ref{fig:central_pressurevsmass},
we show data points for $M-R$ using the same color scheme as in the corresponding left panels.
Interestingly, the minimum pressure does not correspond to the minimum mass in Fig.~\ref{fig:central_pressurevsmass}, but it is an intermediate point in the $M-R$ diagram.
In the center right panel we show an extended version of the $\Lambda=0.141$, $\eta=1.26$ curve shown in Fig.~\ref{fig:massradious}. This highlights the peculiar behavior of the new branch, which has a further turning point at large radii.
Studying the stability of these different peculiar branches~\cite{Guerra:2019srj} is left for future work.\footnote{We point to Ref.~\cite{Mathieu:1983ac}, where a broad class of related theories is analyzed in terms of energy stability (though without taking gravity into account), and to Ref.~\cite{Kusmartsev:1991pm}, in which stability of neutron and boson stars is studied through catastrophe theory.
However, the issue of stability in the present work remains open and needs a full radial perturbation analysis.}

Finally, note that in both cases there are values of the central fermionic pressure corresponding to multiple solutions, each one identified by a different central value of the scalar field.

\section{Parameter space and astrophysical implications}\label{astro_section}
Given the number of parameters of our model, it is interesting to study the characteristic mass and radius of fermion soliton stars in this theory.
By defining
\begin{equation}
    q \equiv (\mu \phi_0^2)^{1/3},
\end{equation}
as long as we are in the confining regime, one finds
\begin{align}\label{eqscalingastro}
		M_c & \sim \frac{0.19}{8\pi} \frac{m_p^4}{q^3} 
		\sim 1.27 \,M_\odot \left( \frac{q}{5 \times 10^5 \, {\rm GeV}}\right )^{-3},
		\\\label{eqscalingastro2}
		R_c &\sim\frac{0.71}{8\pi} \frac{m_p^2}{q^3} 
		\sim  6.5 \, {\rm km} \left( \frac{q}{5 \times  10^5 \, {\rm GeV}}\right )^{-3}, 
\end{align}
where we included the prefactors obtained using the numerical results.
Given the cubic dependence on $q$, the model can accommodate compact objects of vastly different mass scales, while the compactness at the maximum mass is independent of $q$, $GM_c/R_c\sim 0.27$, which is slightly larger than that of a typical neutron star, but still smaller than the compactness of the photon sphere. As a consequence, one expects fermion soliton stars to display a phenomenology more akin to ordinary neutron stars than to black holes~\cite{Cardoso:2019rvt}.
The authors of Ref.~\cite{Lee:1986tr} 
considered the value $q = 30\,  {\rm GeV}$, yielding supermassive objects with
$M_c \sim 10^{12} \, M_\odot$ 
and 
$R_c \sim 10^{13}\,  {\rm km} \sim 0.3\,  {\rm pc}$.
Instead, the choice 
\begin{equation}\label{q_astrophysical}
	q=q_{\rm astro}  
\sim 5 \times 10^5 \, {\rm GeV}
\end{equation}
leads to the existence of soliton solutions of mass and radius comparable to ordinary neutron stars.

Furthermore, the fact that the model is in the confining regime only above a critical value of $\eta$, Eq.~\eqref{condition_on_eta2}, implies (using Eq.~\eqref{dimensionless_parameters} and our numerical results)
\begin{equation}\label{mf_minimum}
	m_f > 2.7 \Big(\frac{\sqrt{8\pi}q^3}{m_p}\Big)^{1/2}\sim 0.6\,{\rm GeV} \left(\frac{q}{q_{\rm astro}}\right)^{3/2},
\end{equation}
a range including the neutron mass. 
Therefore, the fermion gas can be a standard degenerate gas of neutrons.
It is also interesting to combine the above inequality (saturated when $m_f=m_f^{c}$) with Eq.~\eqref{eqscalingastro}, finding a relation between the maximum mass of the soliton in the confining regime and the critical fermion mass,
\begin{equation}
 M_c\sim 0.46 \left(\frac{{\rm GeV}}{m_f^{c}}\right)^2 M_\odot\,,
\end{equation}
independently of $q$.
Interestingly, this model allows for subsolar compact objects for fermions at (or slightly heavier than) the GeV scale, whereas it allows for supermassive ($M_c\sim 10^6 M_\odot$) compact stars for a degenerate gas of electrons ($m_f^{c}\sim 0.5\,{\rm MeV}$).

Clearly, the same value of $q$ can be obtained with different combinations of $\mu$ and $\phi_0$. In general,
\begin{align}\label{eq:muvsq}
 \mu &=500 \left(\frac{q}{q_{\rm astro}}\right)^3 \left(\frac{500\,{\rm TeV}}{\phi_0}\right)^2\,{\rm TeV}\\
 &=500 \left(\frac{m_f^{c}}{0.6\,{\rm GeV}}\right)^2 \left(\frac{500\,{\rm TeV}}{\phi_0}\right)^2\,{\rm TeV}\,,
\end{align}
so $\mu\sim {\rm GeV}$ for $q=q_{\rm astro}$ (or, equivalently, for $m_f^{c}=0.6\,{\rm GeV}$) and $\phi_0\sim3\times 10^5 \,{\rm TeV}$. Note that the latter value is still much smaller than the Planck scale, so the condition $\Lambda\ll1$ is satisfied. 
From our numerical results, Eqs.~\eqref{eqscalingastro} and \eqref{eqscalingastro2} are valid as long as $\Lambda\lesssim 0.5$, whereas, for larger values of $\Lambda$, $M_c$, $R_c$, and $C_{ c}$ decrease rapidly and the condition $\mu R\gg1$ might not hold (see Fig.~\ref{fig:scaling1}). This gives an upper bound on $\phi_0$, 
\begin{equation}
    \phi_0 \lesssim  \frac{0.5}{\sqrt{8\pi}}m_p \sim 10^{18} \, {\rm GeV},
\end{equation}
which, using Eq.~\eqref{eq:muvsq}, can be translated into a lower bound on $\mu$ 
\begin{equation}
   \mu \gtrsim 8.4 \times 10^{-11}  \left(\frac{q}{q_{\rm astro}}\right)^3 \,{\rm eV}.
\end{equation}
Thus, also the scalar-field mass can vastly change depending on the value of $q$, reaching a lower limit that can naturally be in the ultralight regime.

Finally, in the deconfining regime there is no minimum fermion mass so solutions can exist also beyond the range dictated by Eq.~\eqref{mf_minimum}, but soliton fermion stars in such a regime would be characterized by smaller values of the compactness (see discussion in Sec~\ref{sec_num_res}).

\section{Conclusions}\label{sec:conclusions}
We have found that fermion soliton stars exist
as static solutions to Einstein-Klein-Gordon theory with a scalar potential and a Yukawa coupling to a fermion field. This confirms the results of Ref.~\cite{Lee:1986tr} obtained in the thin-wall approximation and provides a way to circumvent the no-go theorems~\cite{Derrick:1964ww,Herdeiro:2019oqp} for solitons obtained with a single real scalar field.

Focusing on spherical symmetry, we have explored the full parameter space of the model and derived both analytical and numerical scalings for some of the relevant quantities such as the critical mass and radius of a fermion soliton star.
Interestingly, the model predicts the existence of compact objects in the subsolar/solar (resp.\ supermassive) range for a standard gas of degenerate neutrons (resp.\ electrons), which might be connected to an exotic explanation for the LIGO-Virgo mass-gap events that do not fit naturally within standard astrophysical scenarios.

We also unveiled the existence of a confining and deconfining regime --~where the macroscopic properties of the soliton are mostly governed by the scalar field parameters or by the fermion mass, respectively~-- and the fact that no Newtonian analog exists for these solutions for fermion masses below a certain threshold.

Extensions of our work are manifold. 
First of all, for simplicity, we have focused on a scalar-fermion coupling tuned to provide an almost vanishing effective fermion mass in the stellar core.
This assumption imposes $f=m_f/\phi_0$, a condition that can be relaxed, thus increasing the dimensionality of the parameter space.
We have also considered a scalar potential with two degenerate minima. A straightforward generalization is to break this degeneracy and allow for a true false-vacuum potential in which the scalar field transits from the false-vacuum state inside the star to the true-vacuum state at infinity.

From the point of view of the fundamental theory, it would be interesting to investigate an embedding within the Standard Model and beyond, also including gauge fields (e.g., see Ref.~\cite{Endo:2022uhe} for a recent attempt along this direction).

Finally, although we focused on static and spherically symmetric solutions, there is no fundamental obstacle in considering spinning configurations and the dynamical regime, both of which would be relevant to study the phenomenology of fermion soliton stars, along the lines of what has been widely studied for boson stars~\cite{Liebling:2012fv} and for mixed fermion-boson stars~\cite{Valdez-Alvarado:2012rct}.
In particular, due to the existence of multiple branches~\cite{Guerra:2019srj} and the absence of a Newtonian limit in certain cases, an interesting study concerns the radial linear stability of these solutions. 

We hope to address these points in future work.

\acknowledgments
We thank Enrico Barausse, Mateja Bošković, and Massimo Vaglio for useful conversations.
G.F. and P.P. acknowledge financial support provided under the European
Union's H2020 ERC, Starting Grant Agreement No.~DarkGRA--757480, under
MIUR PRIN (Grant No. 2020KR4KN2 “String Theory as a bridge between Gauge Theories and Quantum Gravity”) and FARE (GW-NEXT, CUP: B84I20000100001, 2020KR4KN2) programs, and support from the Amaldi Research Center funded by the MIUR program ``Dipartimento di Eccellenza" (CUP:~B81I18001170001). The research of A.U. was supported in part by the MIUR under Contract No. 2017\,FMJFMW (``{New Avenues in Strong Dynamics},'' PRIN\,2017).
This work was supported by the EU Horizon 2020 Research and Innovation Programme under the Marie Sklodowska-Curie Grant Agreement No. 101007855.

\appendix

\section{Connection with scalar-tensor theories}\label{appendix}
In this appendix we discuss whether the model for fermion soliton stars presented in the main text can also arise in the context of a scalar-tensor theory of gravity (see, e.g.,~\cite{Berti:2015itd} for a review on modified theories of gravity).

In the so-called Jordan frame,\footnote{In this appendix we used a hat to denote quantities in the Jordan frame, whereas quantities without the hat refer to the Einstein frame where gravity is minimally coupled to the scalar field.} where gravity is minimally coupled to matter fields, scalar-tensor theories are described by the action (see, for example,~\cite{Sotiriou:2008rp}) 
\begin{align}\label{eq:Jordanframe}
&\hat{S} = \int d^4 x \,
    	\frac{\sqrt{-\hat{g}}}{16\pi G} \Big[ F(\hat{ \phi})\hat{R}  - Z(\hat{ \phi}) \hat{g}^{\mu \nu} \partial_\mu\hat{\phi}\partial_\nu \hat{\phi} - \hat{U}(\hat{\phi}) \Big]
    	\nonumber \\
& +\hat{S}_m (\hat{\psi}_m; \hat{g}_{\mu \nu})\,.
\end{align}
The coupling functions $F$ and $Z$ single out a particular theory within the class. For example, Brans-Dicke theory corresponds to $F = \hat{\phi}$
and $Z = {\omega_0}/{\hat{\phi}}$, where $\omega_0$ is a constant.

We can write the theory in an equivalent form in the so-called Einstein frame, where gravity is minimally coupled to the scalar field. For this purpose, 
we perform a conformal transformation of the metric, $\hat{g}_{\mu \nu} = A^2(\phi) g_{\mu \nu}$ with $A(\phi) = F^{-1/2} (\hat{ \phi})$, a field redefinition, $\phi = \phi(\hat{\phi})$, and a conformal rescaling of the matter field, $\hat{\psi}_m\rightarrow\psi_m$.
The scalar field $\phi$ is now minimally coupled to $g_{\mu \nu}$, whereas 
$\psi_m$ is minimally coupled to $\hat{g}_{\mu \nu}$~\cite{Sotiriou:2008rp}.
The energy-momentum tensor is $T_{\mu\nu} = A^2(\phi)\hat{T}_{\mu\nu}$, whereas the scalar potential becomes $U(\phi) = \frac{\hat{U}(\hat{\phi})}{16\pi G  F^2(\hat{\phi})}$

The scalar field equation in the Einstein frame reads
\begin{equation}\label{eq:scalarfield_EF}
	\Box \phi = -T \frac{d \log A(\phi)}{d\phi} + \frac{\partial U}{\partial \phi}\,.
\end{equation}

Since in our theory~\eqref{theory_fund} the scalar field is minimally coupled to gravity, it is natural to interpret it in the context of the Einstein frame. Thus, we can compare Eq.~\eqref{eq:scalarfield_EF} to the second equation in~\eqref{equations_covariant_form}:
\begin{equation}
	\Box \phi = -fS  + \frac{\partial U}{\partial \phi}\,,
\end{equation}
which, using Eq.~\eqref{identityWPF}, can be written as
\begin{equation}
	\begin{split}
		&\Box \phi = \frac{f}{(m_f - f\phi)} T +  \frac{\partial U}{\partial \phi}\,.
	\end{split}
\end{equation}
Therefore, if we identify
\begin{equation}
	\frac{d \log A(\phi)}{d\phi} =   \frac{-f}{(m_f - f\phi)} =  \frac{1}{\phi-\phi_0} \,,
\end{equation}
the scalar equation of our model is the same as in a scalar-tensor theory with coupling $A(\phi)$ in the Einstein frame. Integrating this equation yields (henceforth assumig $A(0)=1$),
\begin{equation}\label{eq:A_expression}
	A(\phi) = 1 - \frac{\phi}{\phi_0} = \frac{m_\eff}{m_f}\,.
\end{equation}
Interestingly, the matter coupling vanishes when $\phi\approx\phi_0$.

It is left to be checked if the gravitational sector of our model is equivalent to that of a scalar-tensor theory with $A(\phi)$  given by Eq.~\eqref{eq:A_expression}.
Let us consider a degenerate gas of noninteracting fermions with mass $m_f$ in the Jordan frame, with energy-momentum 
\begin{equation}\label{eq:TmunuJordan}
	\hat{T}^{\mu\nu} = (\hat{W}+\hat{P})\hat{u}^{\mu}\hat{v}^\nu + \hat{g}^{\mu\nu} \hat{P}
\end{equation}
where, assuming spherical symmetry,
\begin{equation}\label{eq:integrals_tocompute}
	\begin{split}
		& \hat{W}(\hat\rho) = \frac{2}{(2\pi)^3}\int_{0}^{\hat{k}_F(\hat\rho)} d^3 k \, \sqrt{k^2 + m_f^2}\\
		& \hat{P}(\hat\rho) = \frac{2}{(2\pi)^3}\int_{0}^{\hat{k}_F(\hat\rho)} d^3 k \, \frac{k^2}{3 \sqrt{k^2 +         m_f^2}}\,.
	\end{split}
\end{equation}
In spherical symmetry, since the spacetime has the same form as in Eq.~\eqref{eq:general_spacetime}, it is straightforward to minimize the energy of the fermion
gas at a fixed number of fermions (the calculation is exactly the same as the one done to obtain Eq.~\eqref{eq:kFermi}): 
\begin{equation}
	\hat{k}^2_F = \hat{\omega}^2_F e^{-2\hat{u}} -m_f^2\,.
\end{equation}

It is important to notice that in the standard scalar-tensor theory in the Jordan frame there is no Yukawa interaction; therefore, the fermion particles do not acquire any effective mass.

In the Einstein frame, Eq.~\eqref{eq:TmunuJordan} simply reads
\begin{equation}
	T^{\mu \nu} = (W+P)u^\mu u^\nu + g^{\mu\nu}P\,,
\end{equation}
where $W = A^4(\phi) \hat{W}$ and $P = A^4(\phi) \hat{P}$.
Therefore, also in the Einstein frame we have a perfect fluid in the form of a zero-temperature Fermi gas. Let us now compute the expressions of $W$ and $P$ explicitly.
First of all, from Eq.~\eqref{eq:integrals_tocompute}, following the same computation presented in the main text, we get 
\begin{equation}
	\begin{split}
		&	\hat{W} = \frac{m_{f}^4}{8\pi^2} \Big[\hat{x}\sqrt{1+\hat{x}^2} (1 + 2\hat{x}^2) - \log(\hat{x} + \sqrt{\hat{x}^2 + 1})\Big]\\
		& \hat{P} = \frac{m_{f}^4}{8\pi^2} \Big[\hat{x}\Big(\frac{2}{3}\hat{x}^2 - 1\Big)\sqrt{1 + \hat{x}^2} +  \log(\hat{x} + \sqrt{\hat{x}^2 + 1})\Big]
	\end{split}
\end{equation}
where $\hat{x} = \hat{k}_F / m_{f}$. Since $A(\phi) = m_\eff/m_f$,  
we obtain
\begin{equation}
	\begin{split}
		&	W = \frac{m_\eff^4}{8\pi^2} \Big[\hat{x}\sqrt{1+\hat{x}^2} (1 + 2\hat{x}^2) - \log(\hat{x} + \sqrt{\hat{x}^2 + 1})\Big]\\
		& P = \frac{m_\eff^4}{8\pi^2} \Big[\hat{x}\Big(\frac{2}{3}\hat{x}^2 - 1\Big)\sqrt{1 + \hat{x}^2} +  \log(\hat{x} + \sqrt{\hat{x}^2 + 1})\Big]\,.
	\end{split}
\end{equation}
Note that $W(\hat x)$ and $P(\hat x)$ above implicitly define an equation of state that is exactly equivalent to that obtained from $W$ and $P$ in Eqs.~\eqref{eq:WPSexplicit} and ~\eqref{eq:Pexplicit}. This shows that our model can be interpreted as a scalar-tensor theory in the Einstein frame with coupling to matter given by\footnote{Note that our model and the scalar-tensor theory are not exactly equivalent to each other. Indeed, while in the scalar-tensor theory \emph{any} matter field is universally coupled to $A(\phi)\hat g_{\mu\nu}$, in our model this is the case only for the fermion gas, while any other matter field is minimally coupled to the metric, in agreement with the fact that our model is based on standard Einstein's gravity.} $A(\phi)=m_{\rm eff}/m_f$.

Furthermore, note that the dimensionless quantity 
$\hat{x} ={\hat{k}_F}/{m_f}={k_F}/{m_{\rm eff}}=x$ is invariant under a change from the Jordan to the Einstein frame. Therefore, $W$ and $P$ are exactly those given in Eqs.~\eqref{eq:WPSexplicit} and \eqref{eq:Pexplicit}.

Finally, $\hat{S}$ in the Jordan frame reads 
\begin{align}
    	&\hat{S} = \frac{2}{(2\pi)^3}\int_{0}^{\hat{k}_F} d^3 k \, \frac{m_f}{\sqrt{k^2 + m_f^2}}  \\
    	&=\frac{m_{f}^3}{2\pi^2} \Big[\hat{x}\sqrt{1 + \hat{x}^2} -  \log(\hat{x} + \sqrt{\hat{x}^2 + 1})\Big],
\end{align}
while in the Einstein frame\footnote{The fact that $S = A^3 \hat{S}$ can be derived from the condition $A^4(\phi) \hat{T} = T\Rightarrow A^4(\phi) m_f \hat{S} = m_\eff S$.
}
\begin{equation}
	S = A^3 \hat{S} =  \frac{m_\eff^3}{2\pi^2} \Big[x\sqrt{1 + x^2} -  \log(x + \sqrt{x^2 + 1})\Big]\,,
\end{equation}
since $\hat{x} = x$. Thus, also in this case we obtain the same expression as in Eq.~\eqref{eq:Sexplicit}. 

Having assessed that our model can be interpreted in the context of a scalar-tensor theory, it is interesting to study the latter in the Jordan frame. 
In particular, since
\begin{equation}
	A(\phi) = \frac{1}{\sqrt{F(\hat{\phi})}}\,,
\end{equation}
and $A(\phi)=1-\phi/\phi_0$, the coupling function $F(\hat\phi)$ is \emph{singular} in $\hat\phi(\phi_0)$.
In the language of the scalar-tensor theory, we see that in the core of a fermion soliton star, where $\phi\approx \phi_0$ and matter is almost decoupled in the Einstein frame, the scalar field in the Jordan frame is strongly coupled to gravity.

\appendix

\newpage 
\bibliography{refs}

\end{document}